\documentclass[useAMS,usenatbib]{mnras}

\usepackage{Times}
\usepackage{psfrag}
\usepackage{pspicture}
\usepackage{graphicx}
\usepackage{subfig}
\usepackage{amsmath}
\usepackage{amssymb}
\usepackage{comment}
\usepackage{pdflscape}
 \usepackage[flushleft]{threeparttable}

\title[X-ray and radio properties of LAEs at $\rm z \sim 2-6$]{The X-ray and radio activity of typical and luminous Ly$\bf \alpha$ emitters from $\bf z \sim 2$ to $\bf z \sim 6$: evidence for a diverse, evolving population}
\author[J. Calhau, D. Sobral et al.]{Jo\~ao Calhau$^{1}$\thanks{E-mail: j.calhau@lancaster.ac.uk}, David Sobral$^{1}$, S\'{e}rgio Santos$^{1}$, Jorryt Matthee$^{2}$, Ana Paulino-Afonso$^{1,3}$,\newauthor Andra Stroe$^{4,5}$\thanks{Clay Fellow}, Brooke Simmons$^{1}$, Cassandra Barlow-Hall$^{1}$, Benjamin Adams$^{1}$\\
$^{1}$ Department of Physics, Lancaster University, Lancaster, LA1 4YB, UK \\
$^{2}$ Department of Physics, ETH Z\"{u}rich, Wolfgang-Pauli Strasse 27, 8093 Z\"{u}rich, Switzerland \\
$^{3}$ CENTRA - Centro de Astrof\'{i}sica e Gravita\c{c}\~{a}o, Instituto Superior T\'{e}cnico, Av. Rovisco Pais, 1, 1049-001, Lisboa, Portugal \\
$^{4}$ European Southern Observatory, Karl-Schwarzschild-Str. 2, 85748, Garching, Germany\\
$^{5}$ Center for Astrophysics \textbar  \, Harvard \& Smithsonian, 60 Garden St., Cambridge, MA 02138, USA}

\begin{document}

%\date{Accepted -- . Received --; in original form --}
\pagerange{\pageref{firstpage}--\pageref{lastpage}} \pubyear{2019}
\maketitle

\label{firstpage}
\begin{abstract}
Despite recent progress in understanding Ly$\alpha$ emitters (LAEs), relatively little is known regarding their typical black hole activity across cosmic time. Here, we study the X-ray and radio properties of $\sim$4000 LAEs at $2.2<z<6$ from the SC4K survey in the COSMOS field. We detect 254 ($6.8\% \pm 0.4\%$) LAEs individually in the X-rays (S/N$>$3) with an average luminosity of $\rm 10^{44.31\pm 0.01}\,erg\,s^{-1}$ and average black hole accretion rate (BHAR) of $\rm 0.72 \pm 0.01$\,M$_{\odot}$\,yr$^{-1}$, consistent with moderate to high accreting AGN. We detect 120 sources in deep radio data (radio AGN fraction of $3.2\% \pm 0.3\%$). The global AGN fraction ($\rm 8.6\% \pm 0.4\%$) rises with Ly$\alpha$ luminosity and declines with increasing redshift.
For X-ray-detected LAEs, Ly$\alpha$ luminosities correlate with the BHARs, suggesting that Ly$\alpha$ luminosity becomes a BHAR indicator. Most LAEs ($93.1\% \pm 0.6\%$) at $2<z<6$ have no detectable X-ray emission (BHARs$<0.017$\,M$_{\odot}$ yr$^{-1}$). The median star formation rate (SFR) of star-forming LAEs from Ly$\alpha$ and radio luminosities is $7.6^{+6.6}_{-2.8}$\,M$_{\odot}$\,yr$^{-1}$. The black hole to galaxy growth ratio (BHAR/SFR) for LAEs is $<0.0022$, consistent with typical star forming galaxies and the local BHAR/SFR relation. We conclude that LAEs at $2<z<6$ include two different populations: an AGN population, where Ly$\alpha$ luminosity traces BHAR, and another with low SFRs which remain undetected in even the deepest X-ray stacks but is detected in the radio stacks.
\end{abstract}

\begin{keywords}
galaxies: high-redshift, galaxies: AGN, galaxies: star formation, cosmology: observations, galaxies: evolution, X-rays: galaxies, quasars: supermassive black holes.
\end{keywords}

\section{Introduction}\label{intro}

Several studies have investigated the interplay and evolution of the central supermassive black holes (SMBHs) and their host galaxies. Observations reveal that galaxies were undergoing significantly higher star formation rates (SFRs) in the past, with the star formation rate density reaching a peak around $z\sim 2-3$ \citep[e.g.][]{Lilly1996, Sobral2013a, Khostovan2015}. Such peak seems to be roughly coincident with the highest point in the SMBH activity \citep[black hole accretion rates, BHAR, e.g.][]{Madau2014, Shankar2009, Delvecchio2014,Calhau2017}, suggesting a link between them.

Supermassive black holes can initially emerge from massive black hole seeds formed by the direct collapse of gas clouds \citep[][]{LoebRasio1994} or from the merging of smaller black holes, produced from the first stars, which would then form a population of intermediate mass black holes \citep[][]{MadauRees2001,Mezcua2017,Mezcua2018}. While accretion plays the fundamental role on the growth of black holes \citep[e.g.][]{Volonteri2012}, other studies have considered the hypothesis of coalescences during galaxy mergers \citep[][]{Merritt2005}. Theoretically, simulations have explored the growth of black holes (BHs) driven by galaxy mergers \citep[e.g.][]{DiMatteo2005, Hopkins2005} as well as from accretion processes \citep[e.g.][]{Booth2009, Rosas-Guevara2016, Bower2017}. However, the connection between SFR and BHAR remains unclear and simulations still find different correlations for these two quantities based on the selection methods used for the samples \citep[e.g. ][]{McAlpine2017}.

When studying the relation between SFRs and BHARs, different strategies are employed. One approach is to study samples selected due to their clear active galactic nuclei (AGN) signatures, typically strong X-ray (2-8 keV) emission \citep[e.g.][]{Lutz2010, Harrison2012, Mullaney2012, Stanley2015}.  These samples show varying results depending on the luminosity of the sources, with low luminosity samples (L$\rm _{2-8 \, \rm keV}<10^{44} $ erg s$^{-1}$) showing no correlation between BHAR and SFR \citep[e.g.][]{Mullaney2012b, Stanley2015, Azadi2015}, while high luminosity samples (L$_{2-8 \, \rm keV}>10^{44} $ erg s$^{-1}$) show either  positive \citep[e.g.][]{Lutz2010}, negative \citep[e.g.][]{Page2012} or no correlation at all \citep[e.g.][]{Harrison2012, Stanley2015, Azadi2015}. The differences in the results might be explained by low number statistics and it is worth noting that studies making use of larger samples appear to support the existence of a flat relation between BHAR and SFR at higher X-ray luminosities \citep[e.g.][]{Stanley2015, Azadi2015}. This has been interpreted as, for example, a result of the high variability of AGN activity weakening any observable relation between the SFR and AGN luminosity \citep[e.g.][]{Stanley2015}, or due to an underlying connection such as a common gas reservoir for both AGN and SFR \citep[][]{Azadi2015}. Another explanation might be that, although SFR and BHAR trace each other at the early stages of galaxy evolution, that relation does not continue past a certain point in the galaxy's life. An example of this is the work of \cite{FerreMateu2015}, who found eight massive outlier galaxies in the local M$_{\rm BH}$-M$_{\rm bulge}$ relation \citep[see e.g.][using both early and late-type galaxies]{McConnell2013} and explain them as relics from the $z \sim 2$ Universe whose extremely large SMBHs are completely formed by this redshift and whose host galaxies remain structurally the same and without further growth \citep[see also][]{Barber2016}.

An alternative strategy is to study the BHARs of star forming-selected galaxies (SFGs), which allows for the study of galaxy samples without requiring clear AGN activity. Studies focusing on SFGs consistently find  that the BHAR/SFR ratio stays relatively constant across redshift \citep[e.g.][]{Rafferty2011, Delvecchio2015, Calhau2017}.
A possible explanation is that the relation is due to the BHARs being dependent on the content of dense molecular gas of the host galaxies and, as such, BHARs and SFRs broadly trace each other across cosmic time.

%%%%%%%%%%%%%%%%%%%%%%%%%%%%%%%%%%%%%%%%%%%%%%%%%%%%%
% Figure 1 - Fields
%%%%%%%%%%%%%%%%%%%%%%%%%%%%%%%%%%%%%%%%%%%%%%%%%%%%%
\begin{figure*}
\centering
\includegraphics[width=14cm]{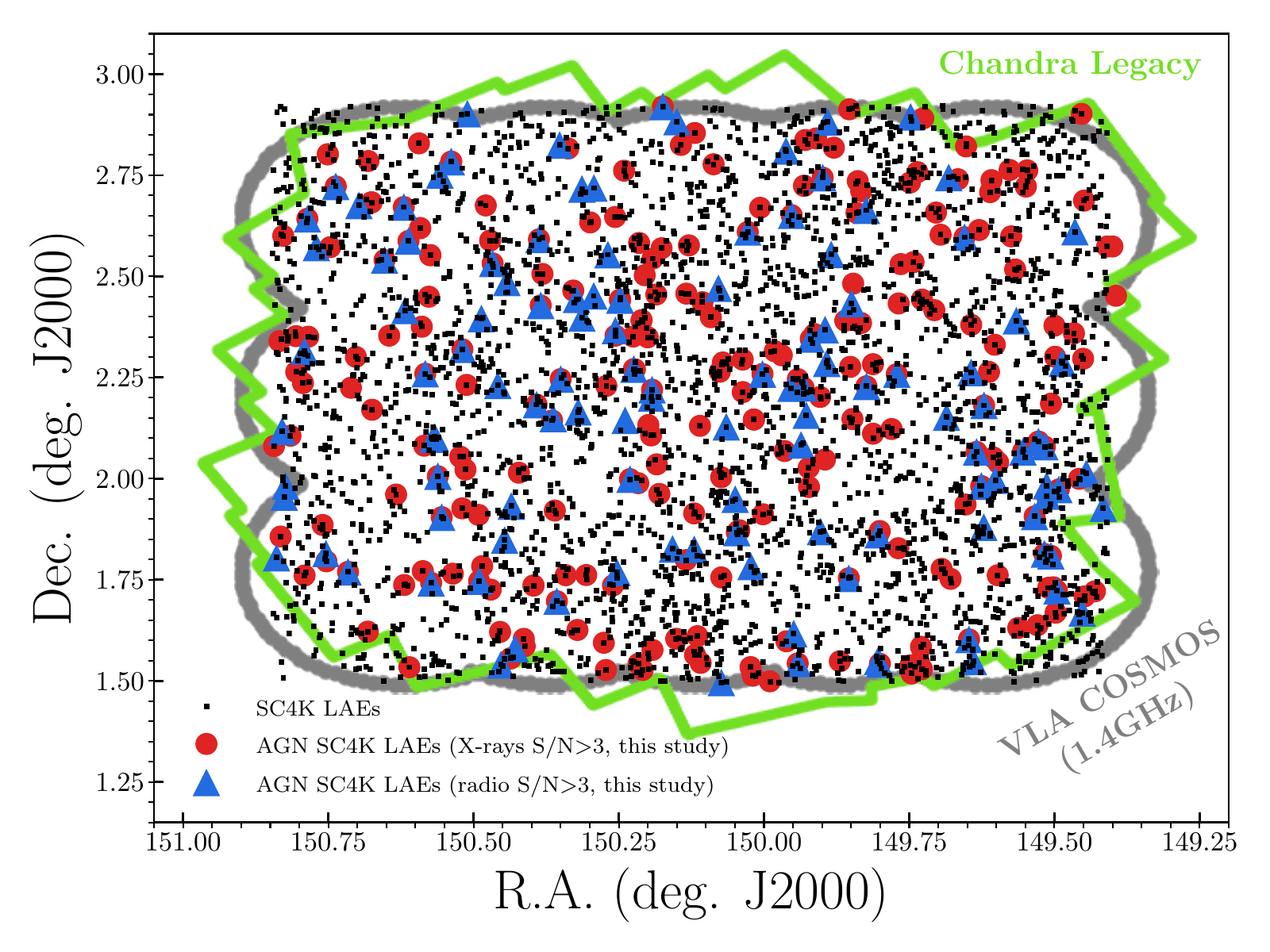}
\caption{The distribution of the SC4K LAEs \citep[][]{SC4Kpaper} across the COSMOS field (black markers). We consider only the sources covered by {\it Chandra} COSMOS Legacy \citep[][green line]{ChandraLegacy}, for a total of 3700 sources. The red circles and blue triangles show the LAEs that are directly detected in the X-rays or radio, respectively. The grey boundary illustrates the area of the VLA-COSMOS 1.4 GHz survey \citep[][]{Schinnerer2004}, which we also use. The HeRMES survey \citep[\textit{Herschel} space telescope, 100, 160, 250, 350 and 500\,$\mu$m -][]{Griffin2010,Oliver2012} and the VLA COSMOS 3 GHz survey \citep[][]{Smolcic2017} cover the totality of SC4K.}
\label{fig:Fields}
\end{figure*}

Further improving our understanding requires large samples of galaxies across cosmic time. With respect to star forming galaxies, we can use the H$\alpha$ ($\rm \lambda_0 = 6563 \, \AA$) emission line to select large and representative samples of SFGs at $z<2.5$ \citep[e.g. the HiZELS survey;][]{Sobral2013a} because it is a very well calibrated star formation indicator with limited dust attenuation and traces SFRs on timescales of $\sim$10 Myr \citep[e.g.][]{Kennicutt1998, Garn2010, Sobral2012, Oteo2015}. 
At $z > 2.5$, H$\alpha$ becomes unobservable from the ground, due to the line shifting into the mid-infrared, but Ly$\alpha$ ($\rm \lambda_0 = 1216 \, \AA$)  may be used as an alternative for tracing both star formation and BH activity \citep[e.g.][]{Ono2012, Stark2015,CALYMHA2017,sobral2019}. It is usually associated with SF  and early ``primeval" galaxies \citep[e.g.][]{Pritchet1994, Cowie1998, Pirzkal2007}, although it can also originate from AGN activity \citep[e.g.][]{Gawiser2006, Ouchi2008, Wold2014, Wold2017, Sobral2018}. Furthermore, at $2<z<7$, the Ly$\alpha$ line is redshifted into the optical band, making it easy to be observed from the ground.

Integral field unit (IFU) and narrow band (NB) surveys have now detected large numbers of Ly$\alpha$ emitters at $2<z<6$, of which several are completely undetected in photometric broad-band surveys. This is consistent with faint LAEs of low mass, blue and with low metallicity \citep[][]{Bacon2015, Karman2015, Sobral2015b,Sobral2019CR7, Nakajima2016}. However, studies at $z\sim2$ find a second population of Ly$\alpha$ sources which are massive, dusty and red \citep[][]{Chapman2005,Oteo2012a, Oteo2015, Sandberg2015, Matthee2016b}.
At lower redshifts ($z<3$) luminous Ly$\alpha$ emitters appear more AGN-dominated \citep[][]{Cowie2010, Wold2014, Sobral2018}, but most may still be considered analogous to $z>3$ LAEs \citep[][]{Oteo2012b, Erb2016, Trainor2016} and the distinction between these two likely depends on the Ly$\alpha$ luminosity, likely due to a maximal observable unobscured Ly$\alpha$ luminosity at $\rm SFR \sim 20 \, M_{\odot}\, yr^{-1}$\citep[e.g.][]{CALYMHA2017,SC4Kpaper}.

Despite evidence that at $z = 2-3$ luminous LAEs (L$_{\rm Ly\alpha} \gtrsim10^{43}$\,erg\,s$^{-1}$) are mostly associated with the presence of AGN \citep[e.g.][]{Ouchi2008,Konno2016, CALYMHA2017}, the limiting X-ray sensitivity makes it impossible to probe the nature of lower luminosity Ly$\alpha$ emitters source by source. \cite{Matthee2017}, for example, reported X-ray fractions as high as $\sim80$\% of luminous Ly$\alpha$ emitters as AGN (L$_{\rm 2-8\,keV}>3 \times 10^{44}$\,erg\,s$^{-1}$ and L$_{\rm Ly\alpha}>10^{44}$\,erg\,s$^{-1}$), while \cite{Sobral2018} showed that such fractions are likely just a lower limit, with the AGN fraction of luminous LAEs being even higher. However, little is known about the potential AGN activity of fainter populations of LAEs. It is uncertain how the AGN fraction of LAEs might evolve with redshift, and whether the transition Ly$\alpha$ luminosity between dominant SF and AGN LAEs evolves with redshift.

In order to make progress we require a large sample of LAEs selected across redshift and with access to the deepest data from X-rays to the radio. In this paper we make use of the public SC4K survey \citep[][]{SC4Kpaper} to study the X-ray and radio properties of roughly 4000 LAEs at $2<z<6$ in the COSMOS field. Using stacking analysis to probe beyond the current limits, we reach an equivalent total exposure time of 482\,Ms ($\sim$15\,yrs) in the X-rays to characterise the AGN activity of LAEs.

This paper is organised as follows: Section \ref{sample_data} describes the sample and data used in this work. Section \ref{Methods} details the methods used to extract the X-ray and radio properties, black hole accretion rates and SFRs. Section \ref{X-ray_results} through \ref{AGN_fraction_results} present the results and discussion on the properties of our sources and their link to Ly$\rm \alpha$ properties. Section  \ref{conclusions} gives the conclusions. In this work, we use a Chabrier initial mass function \citep[]{Chabrier2003} and the following flat cosmology: $\rm H_0=70 \, km\,s\, ^{-1}\,Mpc^{-1}$, $\rm \Omega_M=0.3$ and $\rm \Omega_{\Lambda}=0.7$.

\section{Data and sample}\label{sample_data}

\subsection{The sample of Ly$\alpha$ emitters at $\bf z=2.2-5.8$}\label{Hizels_survey}

We use a large sample of LAEs selected over a redshift range of $z\sim2-6$ in the COSMOS field \citep[SC4K;][]{SC4Kpaper}. SC4K also includes the CALYMHA COSMOS sample at $\rm z = 2.2$ \citep[][]{CALYMHA2017}, with H$\alpha$ coverage from HiZELS \citep[][]{Geach2008,Sobral2009a,Sobral2013a}. The LAEs were detected using a compilation of 16 narrow and medium band (MB) data taken with the Subaru and the Isaac Newton Telescopes. Sources were selected as LAEs through a combination of photometric and spectroscopic redshifts as well as colour-colour diagnostics. Briefly, a LAE satisfies all the following conditions:

\begin{enumerate}
\item Significant excess in a medium (narrow) band, with an EW$_0>50 (25)$\,{\AA} (the majority of LAEs come from MB samples);
\item Presence of a Lyman break in rest-frame wavelengths blue-ward of the identified emission line;
\item A colour cut to exclude dusty lower redshift sources. 
\end{enumerate}

We refer the interested reader to \cite{SC4Kpaper}, for further details on the selection process.
The resulting sample has 3908 LAEs with an average luminosity of $\rm L_{Ly\alpha}\sim 10^{42.9} \, erg \, s^{-1}$ ($\rm \approx L_{Ly\alpha}^*$), over a volume of $\sim$6$\times$10$^{7}$\,Mpc$^3$. We refer to \cite{SC4Kpaper} for the full selection criteria and further details regarding the SC4K LAEs. Further information regarding the rest-frame UV morphologies and sizes of SC4K LAEs can be found in \cite{PaulinoAfonso18} and \cite{Shibuya2019}, while the clustering properties of LAEs and their dependencies on Ly$\alpha$ luminosity and SFRs have been extensively studied by \cite{Khostovan2019}.

Figure \ref{fig:Fields} shows the on-sky distribution of SC4K LAEs in the COSMOS field. We also show the coverage of the {\it Chandra} COSMOS Legacy Survey \citep[][]{ChandraLegacy} and the VLA COSMOS surveys \citep[][]{Schinnerer2004,Smolcic2017}. Note that some SC4K LAEs fall outside the coverage of the {\it Chandra} COSMOS Legacy Survey and we further exclude 5 sources for being too close to the edge of the field, so we use a total of 3700 sources. This constitutes our sample of LAEs.

\subsection{X-ray data: {\it Chandra} COSMOS-Legacy} \label{COSMOS_data}

The \textit{Chandra} COSMOS-Legacy survey \citep[][]{Elvis2009,ChandraLegacy} covers the COSMOS field \citep[e.g.][]{Scoville2007,Capak2007} over a total area of 2.2 deg$^2$. The survey has an exposure time of 150 ks px$^{-1}$ in the central 1.5 deg$^2$ and between 50\,ks\,px$^{-1}$ to 100 ks px$^{-1}$ in the external regions. The average flux limit of the survey, as defined by the source catalogue \citep[][]{ChandraLegacy} is $8.9\times10^{-16}$\,erg\,s$^{-1}$\,cm$^{-2}$ for the full band ($0.5-7$\,keV), $\rm 2.2 \times 10^{-16}$ erg s$^{-1}$ cm$^{-2}$ for the soft band ($\rm 0.5-2$ keV) and $1.5\times10^{-15}$\,erg\,s$^{-1}$\,cm$^{-2}$ for the hard band ($\rm 2-7$ keV).

Figure \ref{fig:Fields} shows an illustration of the regions covered by each of the surveys used in this work and the sources classified as X-ray AGN, in comparison to SC4K. The deep X-ray data allow us to track X-ray emission from processes like Bremsstrahlung and inverse-Compton scattering \footnote{Mainly inverse-Compton scattering, as thermal emission becomes negligible at higher redshifts, \citep[see][]{Lehmer2016}.}, and thus to identify AGN X-ray emission.

\subsection{Radio data: 1.4\,GHz and 3\,GHz VLA-COSMOS}

The VLA-COSMOS Survey \citep[]{Schinnerer2004, Schinnerer2007, Bondi2008, Schinnerer2010} used the National Radio Astronomy Observatory's Very Large Array (VLA) to conduct deep, wide-field imaging with $\approx1.5''$ resolution at 1.4\,GHz continuum of the 2 deg$^2$ COSMOS field (Figure \ref{fig:Fields}). The data reaches down to a 1$\sigma$ sensitivity of about 11\,$\mu$Jy\,beam$^{-1}$, leading to \cite{Bondi2008} presenting a catalog of roughly 3600 radio sources.

The VLA's 3\,GHz COSMOS Large Project covers the entirety of the COSMOS field at a deeper average sensitivity of 2.3\,$\mu$Jy\,beam$^{-1}$ and also at a higher spatial resolution, with an average beam-width of 0.75\,$''$. The observations and data reduction details can be found in \cite{Smolcic2017}, including a catalogue of over 10,000 radio sources. Usage of the 3\,GHz VLA data allows us to further probe the existence of radio-emitting AGN over a larger area, as this survey covers the entirety of SC4K. In addition, by removing radio-detected AGN and obtaining deep radio stacks, radio data will allow us a dust-independent determination of the SFRs of SC4K LAEs (as confirmed in Section  \ref{SFRs}). We nevertheless caution that removing the radio-detected sources may still result in some contamination from undetected low luminosity radio AGN.

\section{Methodology}\label{Methods}

Here we present the full methodology leading to all the quantities that are explored in this paper. These include X-ray, radio, FIR and Ly$\alpha$ derived properties. We make all computed quantities fully available in a new electronic SC4K public catalogue to be published with this paper.

\cite{ChandraLegacy} and \cite{Smolcic2017} provide catalogues with X-ray and radio-detected sources, respectively. We match our sources with the \cite{ChandraLegacy} X-ray and \cite{Smolcic2017} radio catalogues, using 1'' matching radius, and perform visual checks to exclude sources contaminated by nearby X-ray or radio emission. We obtain 100 matched sources for the X-rays and 54 matched sources for the radio, which we use to correct our fluxes in Sections \ref{x-ray_flux} and \ref{radio_alpha}. We also conduct a simpler and uniform source extraction and analysis because, since we have pre-selected sources and know their RA and Dec, we can afford to use a lower signal-to-noise ratio and select X-ray and radio emitters down to $\rm S/N=3$.

\subsection{X-ray analysis}\label{X-rays}

X-rays are one of the most efficient ways to probe the activity of black holes because they track the accretion of matter into the BH directly from the photons emitted through inverse Compton effect on the accretion disk \citep[]{Haardt1991}. Because of this, we expect X-ray luminosity to scale with the BHAR and use it to not only identify AGN, but also to estimate the growth rate of the supermassive black hole.

\subsubsection{Source detection}\label{x-ray_sources}

For our X-ray analysis, we make use of the data from the \textit{Chandra} Legacy Survey \citep[][]{ChandraLegacy}, which builds upon the C-COSMOS survey \citep[][]{Elvis2009}. We bin the original science images (pixel scale of 0.5$''$px$^{-1}$) by a factor of 2 to match the corresponding exposure maps. We obtain cut-outs of 100$\times$100\,px for both the X-ray and exposure maps centred on the 3700 LAEs and mask pixels with 0 exposure times, before transforming the image to counts/s.
In this study we use apertures with a diameter of 8\,px ($\rm \sim7.9''$), centred on the position of each LAE. This aperture allows us to extract roughly the full fluxes of most sources \citep[$\rm \sim80\%$ - see][]{ChandraLegacy} in the COSMOS-Legacy survey without adding significant noise to the measurements. Nevertheless, we apply a final (small) aperture correction to assure we recover the full fluxes (see Section \ref{x-ray_flux}).

\subsubsection{Background and net count estimation}\label{background}

To determine the background counts, we randomly place 7.9$''$ apertures, while ignoring the central area and image borders. We restrict the placement of empty 8\,px apertures to a region of 100$\times$100\,px around each LAE rather than the entire \textit{Chandra} image in order to measure the local noise and background. The counts/s in each of the individual apertures sampling the background are summed and the median of 2000 apertures is taken as the background value for each source. This background value is then subtracted from the source's net count.

The uncertainty is measured by taking asymmetric errors. We define the upper and lower errors as the 84th and 16th percentile of the backgrounds, respectively. The signal to noise ratio (S/N) is defined as the ratio between the net counts/s and the lower error of each image. A source is considered detected if its signal-to-noise rises above or equals 3, but we also define S/N cuts of 5 and compare with the higher significance catalogue provided by \cite{ChandraLegacy}.

\subsubsection{X-ray Flux estimation}\label{x-ray_flux}

We convert our counts/s into flux by using the method detailed in \cite{Elvis2009} and \cite{ChandraLegacy}. To this effect, we multiply our normalised count rates by a conversion factor (CF) and divide the result by a factor of 10$^{11}$:
\begin{equation}
\rm F_{X_0}=(counts/s) \times CF \times 10^{-11} \, (erg\,s^{-1}\,cm^{-2})
\end{equation}

In our study, we take the average of the conversion factors between the two {\it Chandra} COSMOS surveys, C-COSMOS \citep[][]{Elvis2009} and {\it Chandra} Legacy \citep[see][]{ChandraLegacy}, resulting in conversion factors of 0.687, 3.05 and 1.64 for our Soft, Hard and Full band CFs, assuming a photon index $\rm \Gamma=1.4$. The noise values are converted in the same way. Note that the correction factors for C-COSMOS we use are corrected values \citep[see][]{ChandraLegacy}.

We compare our aperture fluxes $\rm F_{X_0}$ with the full fluxes F$_C$ obtained by \cite{ChandraLegacy} by matching them using a 1$''$ matching radius and calculate an aperture correction as the median of the flux difference in log space\footnote{$\rm A_{C} = median[\log_{10}(F_C) - \log_{10}(F_{X_0})]$, see also Figure \ref{fig:FluxesComp}.}. This allows us to define a median aperture correction from our fluxes to full fluxes\footnote{We also do a linear fit to the difference between F$_{X_0}$ and F$_C$, before correcting our fluxes, and find that the slope is close to zero (-0.01).}. We define our full flux, $\rm F_X$, aperture corrected to match \cite{ChandraLegacy}, as:
\begin{equation}
\rm \log_{10}(F_X)=\log_{10}(F_{X_0})+A_{\rm C}
\end{equation}
We find $\rm A_{C}=0.1$, which we apply throughout this paper.
The matched \cite{ChandraLegacy} fluxes show a median error of $\rm 8.1\times 10^{-16} \, erg \, s^{-1}$, in the full band, and the individual errors vary across the Chandra Legacy field, with no apparent trend. Our aperture-corrected flux errors for the matched sources, taken as the 84th and 16th percentile of the background measurements, have a median of $\rm 3.1 \times 10^{-16} \, erg \, s^{-1}$ but get greater as the sources approach the edges of the Chandra Legacy field, following the trend set by the exposure maps and in some of the more extreme cases reaching errors of the order of $\rm 10^{-15}  \, erg \, s^{-1}$.

\subsubsection{Hardness ratio estimation}\label{HR}

\begin{figure*}
\centering
\includegraphics[width=17.8cm]{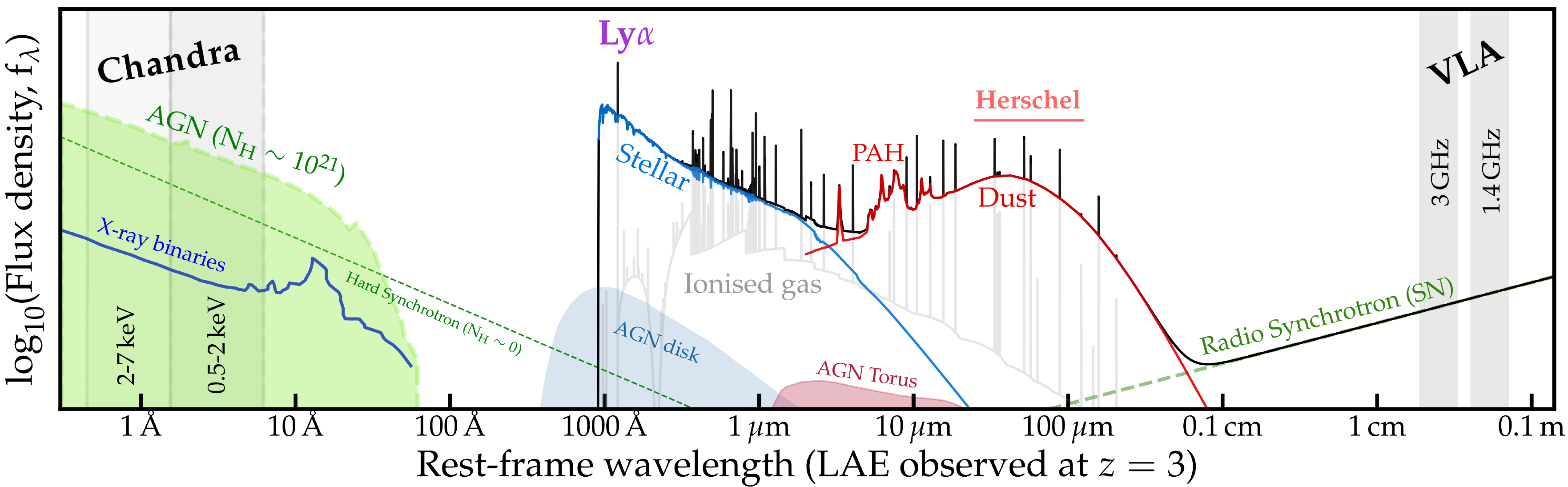}
\caption{An overview of the data used in this study and the mechanisms that originate emission from X-rays through to the radio. We show the X-ray band, divided in its hard (2-7\,keV) and soft (0.5-2\,keV) bands, as well as the expected SED for an AGN emitting in the X-rays through a column density of $\rm N_H\sim10^{21}$\,cm$^{-2}$ \citep[see][]{Hickox2018}. We also include an illustrative contribution from X-ray binaries and the spectrum of a hard synchrotron AGN if subjected to no absorption. Also seen are the contributions by the stellar and nebular components, as well as the AGN accretion disk and ionised gas. We further show the thermal emission due to dust and the contribution in the radio due to synchrotron from radio AGN and supernovae (tracing SFR).}
\label{fig:LAESED}
\end{figure*}

Soft band photons are more efficiently absorbed by the environment surrounding the SMBH than harder photons. We illustrate this in Figure \ref{fig:LAESED}, where we show the spectrum of an AGN observed through a column density of N$_{\rm H}\sim10^{21}$\,cm$^{-2}$ and the emission of hard X-ray photons, from inverse Compton scattering or synchrotron radiation, with no absorption. As the column density increases, the spectrum of the AGN gets increasingly absorbed, starting with the photons from the soft band (0.5-2 keV), which translates into a lower count rate for this band. Comparing the count rates of both the soft and the hard band gives us a measurement of the level of obscuration of an AGN \citep[see, e.g.,][]{Park2006}. We achieve this by estimating the hardness ratio (HR) of the AGN. In this endeavour we restrict ourselves to sources detected in both the soft and hard band.
In order to estimate the hardness ratio of our sources, we adopt the standard definition \citep[see][for a discussion on the various definitions of the hardness ratio]{Park2006}:
\begin{equation}
HR = \frac{H-S}{H+S}
\end{equation}
where H and S are the count rates (counts/s, in 7.9\,$''$ apertures) in the hard (2-7 keV) and soft band (0.5-2 keV), respectively. We caution that the requirement of both soft and hard band detections for the determination of the hardness ratio may bias us towards more obscured sources, in a low count scenario.

%%%%%%%%%%%%%%%%%%%%%%%%%%%%%%%%%%%%%%%%%%%%%%%%%%%%%
% Figure 2 - SEDs
%%%%%%%%%%%%%%%%%%%%%%%%%%%%%%%%%%%%%%%%%%%%%%%%%%%%%

\subsubsection{X-ray luminosity estimation}\label{x-ray_luminosity}
We convert the fluxes to observed X-ray luminosity by using

\begin{equation}
\rm L_X = 4 \pi (F_{X}) {d_L}^{2} \, (erg\,s^{-1})
\end{equation}
where d$_L$ is the luminosity distance in cm. We determine the luminosity distance by taking the redshift associated with the narrow or medium band filter the source is detected with. We do this for both the individual sources and while stacking.

We convert the observed luminosity in each band into the rest-frame $\rm 0.5-10 \, keV$ luminosity by multiplying the observed luminosity by a K-correction factor as defined in \cite{Marchesi2016}, resulting in the expression:
\begin{equation}
\rm L_{0.5-10 \, keV}=\frac{L_X(10^{(2-\Gamma)}-0.5^{(2-\Gamma)})}{(E_{max}(1+z))^{(2-\Gamma)}-(E_{min}(1+z))^{(2-\Gamma)})}
\end{equation}
where E$\rm _{max}$ and E$\rm _{min}$ are the maximum and minimum energies for the band used, $z$ is the redshift and $\Gamma$ is the photon index, assumed to be 1.4. This is the value for the background X-ray slope and is a good average slope for populations containing both obscured and unobscured AGN \citep[assuming Galactic absorption, see][]{Markevitch2003}. It is also a good value for star-forming galaxies \citep[not expected to have strong X-ray emission, see][]{Alexander2003}. We do not correct for absorption at the source since the vast majority of SC4K LAEs are not detected in the X-rays and we have no way of determining their intrinsic absorption. For the sources that are detected in both bands we estimate an average $\rm HR \sim -0.1$, which translates into an absorption of $\rm 1.7 \times 10^{23} \, cm^{-2}$ and a correction of 0.7 to the full band X-ray log scale luminosity. We therefore caution that some X-ray luminosities and BHAR may be underestimated due to their sources being obscured.

\subsubsection{Black Hole Accretion Rates}\label{BHAR_estimation}

In order to determine the BHARs of our sources, we start by translating our $0.5-10$\,keV luminosities into bolometric luminosities by taking:
\begin{equation}
\rm L_{bol}=22.4 \times L_{0.5-10\,keV}
\end{equation}

Where 22.4 is the bolometric correction factor. \cite{Vasudevan2007} find that the bolometric correction varies with the Eddington ratio of the sources, going from $\rm 15-25$ for AGN with Eddington ratios of $<$0.1 and $\rm 40-70$ for AGN with higher ratios. Given the high variability of the bolometric corrections, we follow \cite{Lehmer2013} and assume the median value of 22.4 for the bolometric correction of AGN of $\rm L_X=10^{41}-10^{46} \, erg \, s^{-1}$.
We then estimate the BHAR from our bolometric luminosities using:
\begin{equation}
\rm \dot{M}_{\rm BH}=\frac{L_{\rm bol}(1-\epsilon)}{\epsilon c^2} \times 1.59 \times 10^{-26}\, ( \rm{M_{\odot}\,yr^{-1}})
\end{equation}
where $\rm \dot{M}_{\rm BH}$ is the BHAR, $\epsilon$ is the accretion efficiency, assumed to be 0.1 \citep[see][for motivation]{Marconi2004} and c is the speed of light. We stress that we are assuming a median value of 22.4 for the bolometric correction, but the actual value is uncertain and may vary. Varying the bolometric correction between 15 and 50 results in an uncertainty of the BHAR of the order of $\rm ^{+0.5}_{-0.03} \,M_{\odot}\,yr^{-1}$ and adding it in quadrature would provide a more conservative error estimation. However we adopt the median value for simplicity.

\subsubsection{Stacking}\label{X-ray_stacking}

Apart from studying the individual sources, we also obtain stacks of LAEs in the X-rays. In order to do so we first stack the cut-outs in count/s, using median and average statistics, before following the same procedure as for the individual sources. This include applying the correction to the fluxes estimated from our comparison with \cite[][see Section \ref{x-ray_flux}]{ChandraLegacy}. This means if we stack the individual detections we recover the average (median) fluxes of \cite{ChandraLegacy} catalogue within 0.01 (0.03) dex. We calculate the median of the redshifts of the sources used in the stacks and take it as the redshift associated with that stack, effectively treating all sources in a stack as having that same redshift. We also take the 16th and 84th percentiles as the errors associated with the redshift, where applicable. The median redshifts are then used to estimate the luminosity distances used when calculating the X-ray luminosities of the stacks.

We stack our sources based on different redshift and Ly$\alpha$ luminosity bins (see Table \ref{table:appendix_stacks}). We also stack the full sample, both while including and excluding the AGN candidates.

\subsection{Radio analysis}\label{radio}

\subsubsection{Source detection: 3\,GHz}\label{radio_sources_3}

For our analysis, we use the VLA-COSMOS 3\,GHz Large Project data in Jy/beam. \cite{Smolcic2017} estimate the fluxes of the sources by selecting all pixels above a S/N threshold ($\geq$5) and enforcing a minimum area of 3\,px by 3\,px. The total flux density is taken as the sum of all the values within the area and then dividing it by the beam size in pixels. The peak flux is estimated by fitting a two-dimensional parabola around the brightest pixel. We use a simpler method and fix an aperture with a radius of 0.6$''$ (0.8$\times$beam radius) for a total integration area of 1.76$''^2$. We also apply an aperture correction to recover the \cite{Smolcic2017} fluxes on average (see Section \ref{radio_alpha} and Figure \ref{fig:radioFluxesdelta}).

\subsubsection{Source detection: 1.4\,GHz}\label{radio_sources_14}

For the 1.4\,GHz VLA-COSMOS data, \cite{Schinnerer2007} use AIPS \citep[Astronomical Image Processing System,][]{Greisen2003} to find sources with peak fluxes higher than a certain flux level (30$\rm \mu Jy\, beam^{-1}$, $\rm \sim 3\sigma$ in the most sensitive regions). For each component, AIPS gives the peak flux and total flux, among other quantities, by either Gaussian fitting or applying non-parametric interpolation. The original catalogue has since been updated by \cite{Bondi2008}, which we use in this work. Because the resolution of the 1.4\,GHz band is poorer than 3.0\,GHz, we use a larger fixed aperture of 2.5$''$ radius (1.4$\times$beam radius) for our analysis, for a total integration area of 19.6$''^2$, selected in order to make our fluxes as similar to \cite{Bondi2008} as possible.

\subsubsection{Background estimation}

To determine the background we place 1.2$''$ apertures (0.6$''$ radius) for the 3\,GHz band and 5$''$ apertures (2.5$''$ radius) for the 1.4\,GHz band, masking the image borders and the area centred on the LAE for which we are performing the flux measurement. The placement of the empty apertures is restricted to a region of 100$\times$100\,px around each LAE, allowing us to measure the local background and noise levels. The fluxes in the background areas are summed and the total background is taken as the median of $\sim$2000 random apertures. We then subtract this value from the source's flux.

The uncertainty is taken as the 84th and 16th percentile of the background (upper and lower errors, respectively). We define the S/N as the ratio between the source's flux and the lower error of each image. A source is considered detected if the S/N rises above or equals 3, but we also define S/N cuts of 5 and use the \cite{Smolcic2017} catalogue.

\subsubsection{Radio flux and spectral index estimation}\label{radio_alpha}

We compare our aperture fluxes F$_{\rm \nu_0}$ with the appropriate full fluxes F$_{\rm r}$ obtained by either \cite{Smolcic2017} for 3\,GHz or \cite{Bondi2008} for 1.4\,GHz and calculate an aperture correction (A$_{\rm C}$) per band as the median of the flux difference in log space\footnote{$\rm A_{C} = median[\log_{10}(F_ r) - \log_{10}(F_{\nu_0})]$.}. We define our full radio fluxes (F$_{\rm \nu}$), aperture corrected to median match \cite{Smolcic2017} or \cite{Bondi2008} catalogue fluxes, as:
\begin{equation}
\rm \log_{10}(F_{\rm \nu}) = \log_{10}(F_{\rm \nu_0})+A_{\rm C}
\end{equation}
We find $\rm A_{C}=-0.05$ and $\rm A_{C}=0$ for the 3.0\,GHz and 1.4\,GHz bands, respectively, which we apply throughout this paper. By using the radio fluxes, we also calculate the radio spectral index $\alpha$, estimated between 1.4\,GHz and 3.0\,GHz, as:
\begin{equation}
\rm \alpha = \frac{\log_{10}\left(\frac{F_{\rm 3\,GHz}}{F_{\rm 1.4\,GHz}}\right)}{\log_{10}\left(\frac{3}{1.4}\right)}
\end{equation}
where F$_\nu$ is the flux at frequency $\nu$.

\subsubsection{Radio luminosity estimation}\label{radio_luminosity}

We estimate the radio luminosities by using:
\begin{equation}
\rm L_{\nu}=\frac{4 \pi {d_L}^2}{(1+z)^{\alpha+1}}  F_{\nu} \, (W\,Hz^{-1}),
\end{equation}
where d$_L$ is the luminosity distance in meters, $\rm z$ is the redshift, $\rm F_{\nu}$ is the flux at 1.4\,GHz or 3\,GHz (W\,Hz$^{-1}$\,m$^{-2}$) and $\alpha$ is the radio spectral index. We assume $\alpha = -0.8$, the characteristic spectral index of synchrotron radiation and a value typically found in AGN \citep[][although for a wider redshift range of $0<z<5$]{Delhaize2017}, even though we note that, on average, our sources detected in both 1.4\,GHz and 3\,GHz show a steeper $\alpha$ ($\approx-1.3$). Because the 3\,GHz data is deeper than the 1.4\,GHz, it is possible the steeper indices are a selection effect of the increased depth of the 3\,GHz band. Furthermore, a very steep spectral index may lead to source being more easily detectable in 1.4\,GHz (see Figure \ref{fig:LAESED}). However, the unique advantage of the current deeper 3\,GHz data is the much higher spatial resolution, diminishing the risk of contamination by nearby sources. Therefore, throughout this paper, we chose to make use of the 3\,GHz data whenever possible. We then convert the 3\,GHz fluxes into 1.4\,GHz luminosity by following the steps detailed in \cite{Delhaize2017}:
\begin{equation}
\rm L_{1.4 \, GHz}=\frac{4 \pi {d_L}^2}{(1+z)^{\alpha+1}} \left(\frac{1.4\,GHz}{3.0\,GHz}\right)^{\alpha} F_{3 \, GHz} \, (W\,Hz^{-1})
\end{equation}
where F$\rm_{3 \, GHz}$ is the flux in the 3\,GHz band ($\rm W\,Hz^{-1}\,m^{-2}$), D$\rm _L$ is the luminosity distance in m and $\alpha$ is the spectral index, assumed to be $-0.8$. When referring to luminosities in the radio band, we use the converted 3.0\,GHz$\rightarrow$1.4\,GHz luminosity (hereafter L$_{\rm radio}$) and only use the 1.4\,GHz measurements for the sources that are not detected in the 3\,GHz band (which we specifically refer to as L$_{\rm 1.4\,GHz}$).

\subsubsection{Radio stacking}\label{radiostack}

We perform mean and median stacking in each individual band, both when including all sources and after removing the radio detections. 
As with the X-ray data, we perform stacking for various sub-samples (see table \ref{table:appendix_stacks_radio}), taking the median redshift of each stack in order to calculate the luminosity distances required for estimating radio luminosities.

\subsection{SFRs of LAEs}\label{SFRs}

\subsubsection{FIR SFRs and upper limits}\label{FIR_SFR}

We explore the public \cite{Shuowen2018} COSMOS catalogue with de-blended FIR photometry (100, 160, 250, 350 and 500$\mu$m, with 1$''$ matching radius) in order to obtain the fluxes for the LAEs detected in the FIR \citep[$\rm S/N\geq3$, see further details in][]{Santos2020}. We use the fluxes obtainned from the matching and fit them for each of the 46 LAEs detected in at least one FIR band. We explore a range of modified black-bodies \citep[grey-bodies; see details in e.g.][]{Calhau2017}. We then integrate between rest-frame 100$\mu$m and 850$\mu$m to obtain the total FIR luminosity after using the corresponding luminosity distance of each LAE. We convert the luminosity to SFRs by using:
\begin{equation}
\rm SFR_{IR} = L_{IR} \times 2.5 \times 10^{-44} (M_{\odot} \, yr^{-1})
\end{equation}
where L$\rm _{IR}$ is the luminosity obtained from integrating the grey-body templates \citep[e.g. ][converted to Chabrier IMF]{Ibar2013}. We also compare our SFRs and FIR luminosities with those presented in \cite{Shuowen2018}, finding a good agreement on average and within the errors, with differences mostly arising from sources with just one FIR detection per source.

\subsubsection{Radio SFRs}\label{radio_SFR}

We determine the radio SFRs from the 1.4\,GHz luminosities by adopting the calibration used by \cite{Yun2001}, converted to a Chabrier IMF \citep[see][]{Karim2011}:
\begin{equation}
\rm SFR_{1.4 \, GHz}=3.18 \times 10^{-22} L_{1.4 \, GHz} \, \, (\rm{M_{\odot}\,yr^{-1}})
\end{equation}

We only estimate radio SFRs for the stacks where LAEs directly detected in the radio are excluded, in order to avoid contamination by radio AGN emission.

\subsubsection{Ly$\alpha$ SFRs}\label{Line_SFRs}

We also estimate the SFRs of LAEs by following \cite{sobral2019} and using a Chabrier IMF:

 %%%%%%%%%%%%%%%%%%%%%%%%%%%%%%%%%%%%%%%%%%%%%%%%%%%%%
% Figure 3 - AGN examples
%%%%%%%%%%%%%%%%%%%%%%%%%%%%%%%%%%%%%%%%%%%%%%%%%%%%%
\begin{figure}
\centering
\includegraphics[width=8.4cm]{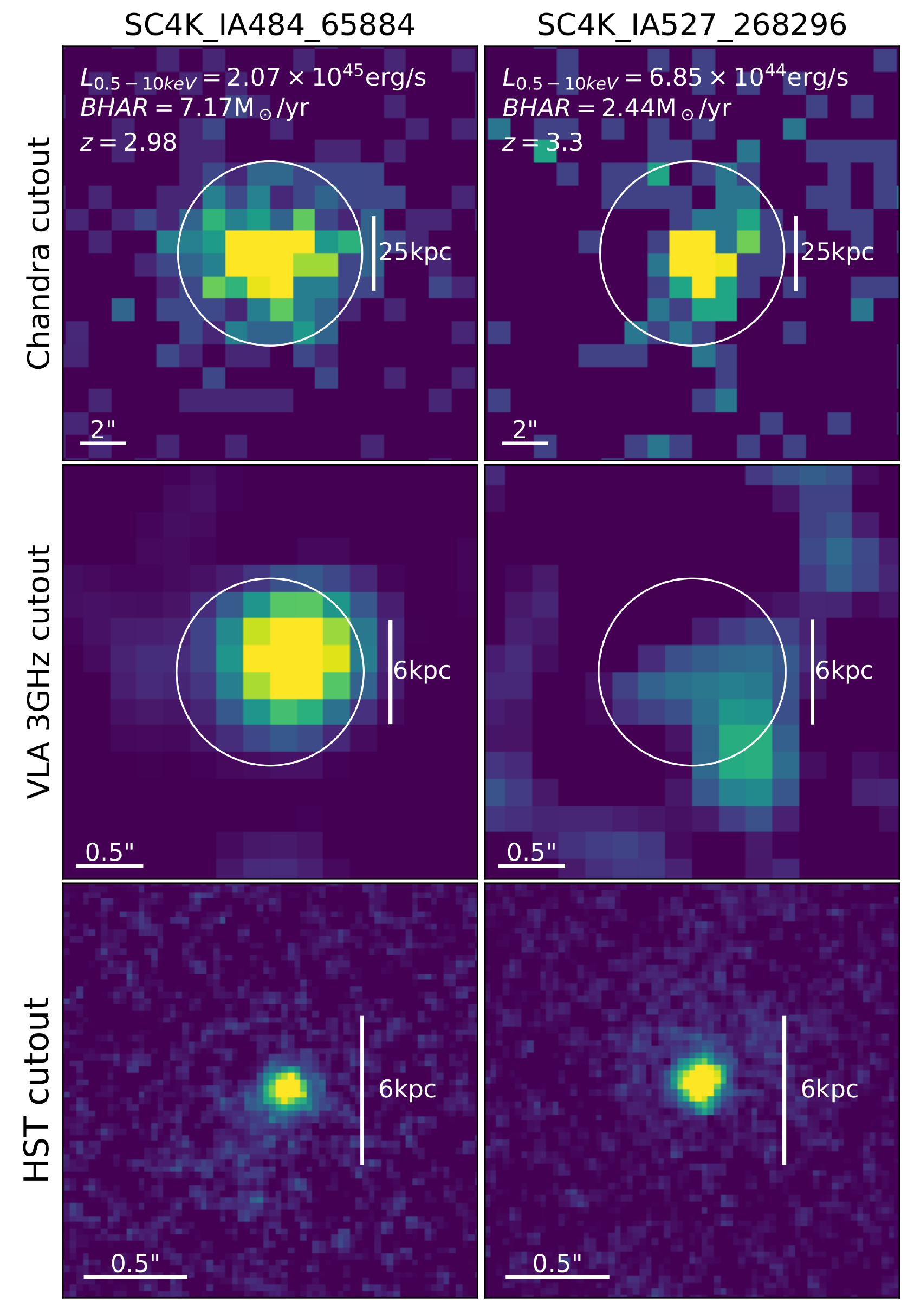}
\caption{Two X-ray detected LAEs: SC4K-IA484-65884 (left, $z=2.98$) and SC4K-IA484-268296 (right, $z=3.3$). The two sources have high X-ray luminosities, implying BHARs of $\sim7$\,M$_{\odot}$\,yr$^{-1}$ and $\sim2$\,M$_{\odot}$\,yr$^{-1}$, respectively. Both present point-like X-ray emission. The circles represent the apertures used for determining the fluxes. The second row of images shows the VLA 3\,GHz cut-outs for the sources, showing that only one of these LAEs is significantly detected in the radio ($\rm S/N\geq3$). The third row shows the {\it HST} F814W filter cut-outs for the respective sources, revealing very compact rest-frame UV morphologies \citep[see also][]{PaulinoAfonso18}.}
\label{fig:AGN}
\end{figure}
%Needs more contrast. 
%Re: Upped contrast. Also centred HST and equalised radio and HST scale.

%%%%%%%%%%%%%%%%%%%%%%%%%%%%%%%%%%%%%%%%%%%%%%%%%%%%%
% Figure 4 - Stacks
%%%%%%%%%%%%%%%%%%%%%%%%%%%%%%%%%%%%%%%%%%%%%%%%%%%%%
\begin{figure}
\centering
\includegraphics[width=8.4cm]{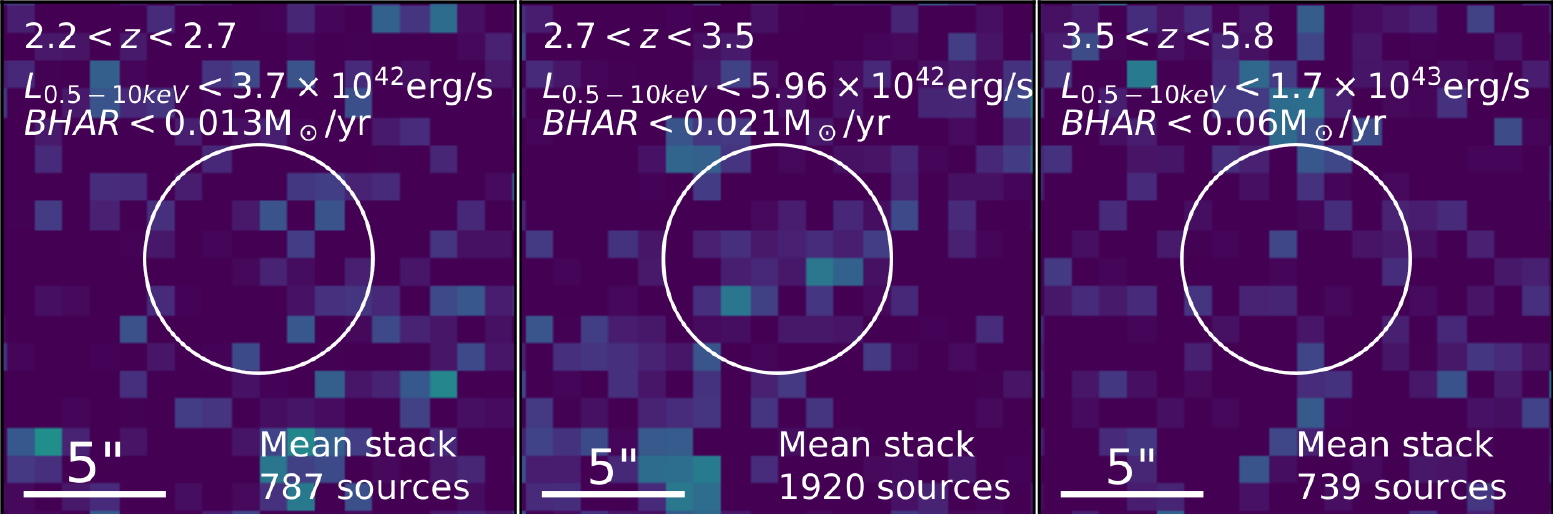}
\caption{The results of mean stacking LAEs in the X-rays in 3 different redshift groups, excluding all LAEs that are individually detected in the X-rays (S/N\,$>3$). No X-ray emission is detected in any of these stacks and we are only able to provide upper limits for the luminosity and BHARs.}
\label{fig:Stacks}
\end{figure}

%%%%%%%%%

\begin{equation}
\rm SFR_{Ly\alpha}[M_{\odot}\,yr^{-1}] = 
	\begin{cases}
		\rm \frac{L_{Ly\alpha} \times 4.4 \times 10^{-42}}{0.042EW_0} & \rm EW_0<210\AA \\
		\rm 4.98 \times 10^{-43} \times L_{Ly\alpha} & \rm EW_0>210\AA
	\end{cases}
\end{equation}
where L$\rm _{Ly\alpha}$ is the observed Ly$\alpha$ luminosity in erg\,s$^{-1}$ and EW$_0$ is the Ly$\alpha$ rest-frame equivalent width in \AA. The SFRs obtained this way should already be corrected for dust extinction as part of the calibration \citep[see][for a detailed explanation]{sobral2019}.

%%%%%%%%%%%%%%%%%%%%%%%%%%%%%%%%%%%%%%%%%%%%%%%%%%%%%
% Figure 5 - HR_vs_redshift
%%%%%%%%%%%%%%%%%%%%%%%%%%%%%%%%%%%%%%%%%%%%%%%%%%%%%
\begin{figure*}
\centering
\includegraphics[width=18cm]{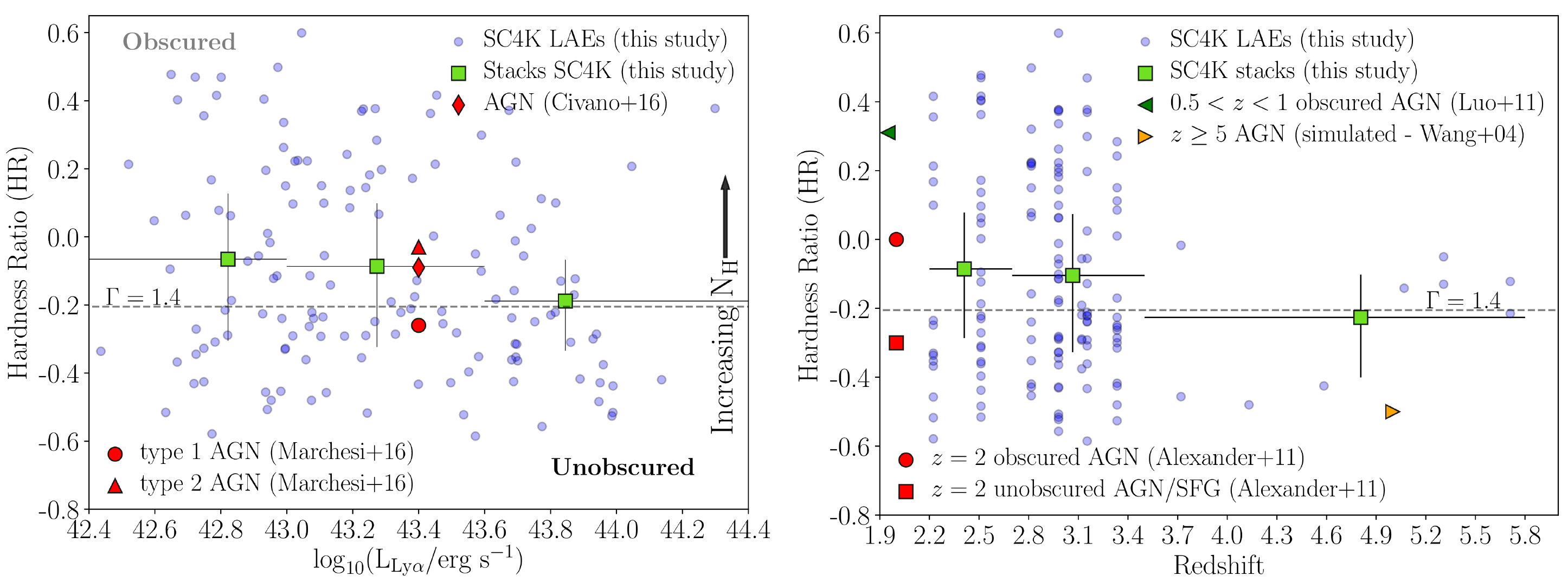}
\caption{The X-ray hardness ratio (HR) of our X-ray AGN LAEs. For our analysis we only use sources detected in both the soft and hard band (S/N\,$>3$), for a total of 143 LAEs. We find no significant relation between the HR and Ly$\alpha$ luminosity or redshift, with roughly half the sources having HR\,$>-0.2$ and therefore consistent with being significantly obscured. The grey line represents the typical HR limit for obscured AGN with $\rm \Gamma = 1.4$ \citep[see also][]{Mezcua2018}. Also shown are the HRs for the samples of \citet{ChandraLegacy} and \citet{Marchesi2016IRChandra} (Left panel) and \citet{Wang2004b} ($\rm z \geq 5$ AGN), \citet{Alexander2011} (obscured AGN and unobscured AGN/SFG) and \citet{Luo2011} (obscured AGN - Right panel). We place literature measurements at their reported redshifts and arbitrarily place measurements at a Ly$\alpha$ luminosity of $\sim10^{43.4}$\,erg\,s$^{-1}$ for illustrative purposes. Our results reveal that X-ray LAEs seem to be representative of the full X-ray selected population.}
\label{fig:HR_z_Lya}
\end{figure*}

%%%%%%%%%%%%%%%%%%%%%%%%%%%%%%%%%%%%%%%%%%%%%%%%%%%%%
% Figure 6 - HR_vs_X-rays
%%%%%%%%%%%%%%%%%%%%%%%%%%%%%%%%%%%%%%%%%%%%%%%%%%%%%
\begin{figure}
\centering
\includegraphics[width=8.4cm]{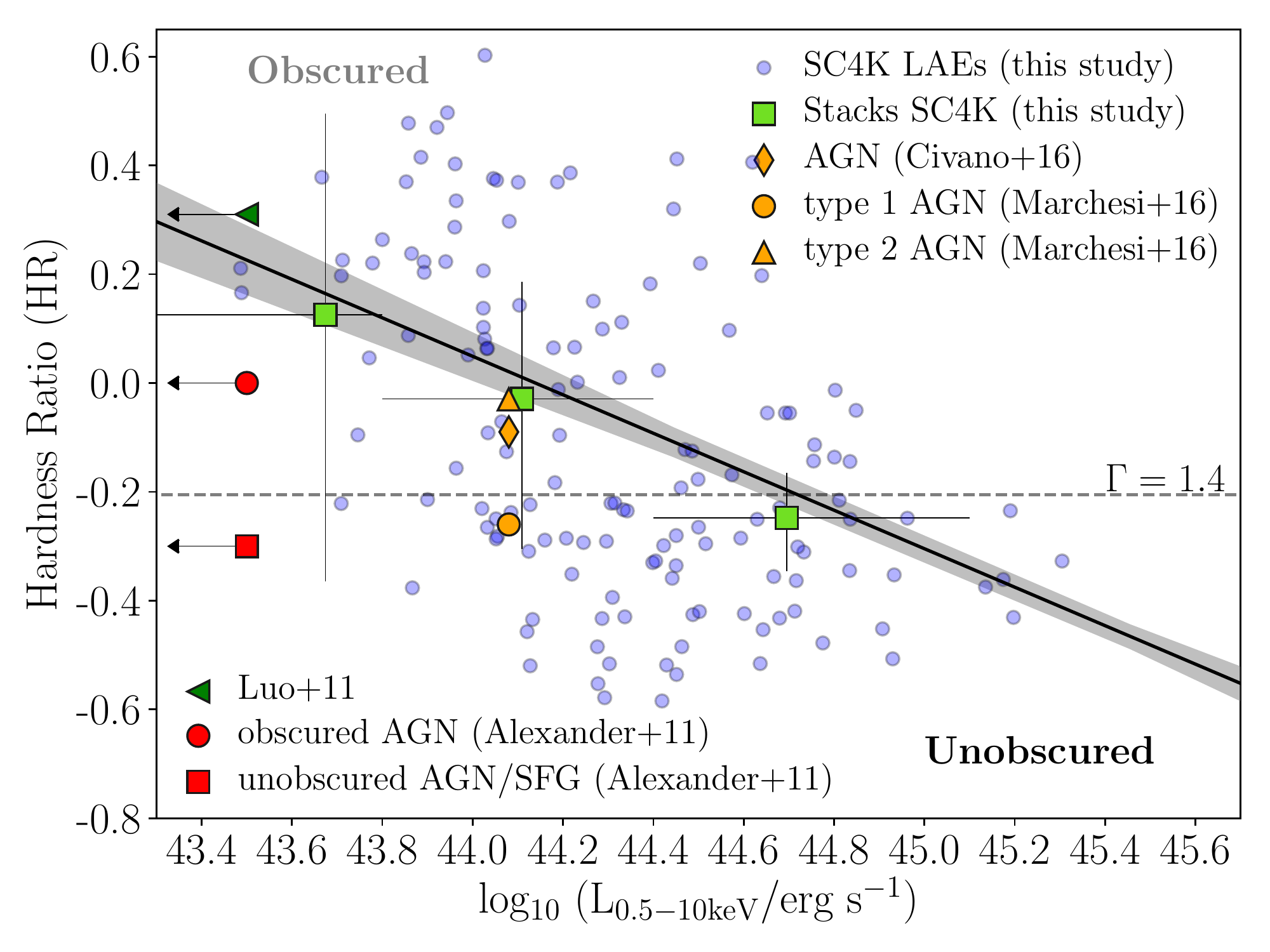}
\caption{The X-ray hardness ratio (HR) of our X-ray AGN LAEs as a function of their X-ray luminosity. A statistically significant correlation is observed, with the more luminous X-ray LAEs having lower HR. This can be interpreted as lower column densities for the LAEs with the highest observed X-ray luminosities, while the lowest X-ray luminosity sources seem to be predominantly highly obscured. For comparison, we show the results by \citet{Alexander2011} (obscured AGN and unobscured AGN/SFG) and \citet{Luo2011} (obscured AGN) for different X-ray luminosities.}
\label{fig:HR_Xrays}
\end{figure}

\subsubsection{Estimating the Ly$\alpha$ escape fraction from EW$_0$ and radio}

We follow \cite{sobral2019} and estimate the escape fraction of Ly$\alpha$ photons ($\rm f_{esc}$) from the median Ly$\alpha$ EW$\rm _0$, by using:
\begin{equation}
\label{eqn:EW_fesc}
\rm f_{esc} = 0.0048 \times EW_{0}
\end{equation}
where EW$\rm _0$ is the median Ly$\alpha$ equivalent width \citep[see full details and the physical interpretation in][]{sobral2019}. With this method we obtain $\rm f_{esc}=0.67^{+0.33}_{-0.34}$ for the full SC4K sample with a median EW$\rm _0$ of 138$^{+282}_{-70}$\,{\AA}.

Using our radio SFRs, we can also estimate the Ly$\alpha$ $\rm f_{esc}$ by simply assuming that $\rm SFR_{1.4\,GHz}=SFR_{H\alpha}$ (see Section \ref{esc_frac_res} for results and discussion), which leads to:
\begin{equation}
\label{eqn:SFR_fesc}
\rm f_{esc}= \frac{L_{Ly\alpha} \times 4.4 \times 10^{-42}}{8.7 \times SFR_{1.4\,GHz}}
\end{equation}
where L$\rm _{Ly\alpha}$ is the observed Ly$\alpha$ luminosity in erg\,s$\rm ^{-1}$ and SFR$\rm _{1.4\,GHz}$ is the radio star formation rate in M$\rm _{\odot}\,yr^{-1}$.

\subsubsection{SFR contribution to the X-ray emission}\label{XSFR}

X-rays can also track SFR due to emission from supernova remnants, high mass X-ray binaries (see e.g. Figure \ref{fig:LAESED}) and hot plasma \citep[see, e.g.][]{Fabbiano1989, Ranalli2003}. We estimate X-ray luminosity (L$_{\rm X}$; erg\,s$^{-1}$) produced by several processes associated with star formation by using the relation\footnote{The relation is given by $\rm log_{10}(L_X) = A + B\,log_{10}(SFR) + C\,log_{10}(1+z)$, and we use A, B and C with values $\rm 39.82 \pm 0.05$, $\rm 0.63 \pm 0.04$ and $\rm 1.31 \pm 0.11$.} derived by \cite{Lehmer2016} for SFGs galaxies at $0<z<7$ (converted to a Chabrier IMF):
\begin{equation}
\rm \log_{10}(L_X) = 39.8 + 0.63\log_{10}(SFR) + 1.31\log_{10}(1+z) \label{Eq_SFR_X}
\end{equation}
where SFR is in M$_{\odot}$\,yr$^{-1}$ and $\rm L_X$ is rest-framed and in erg s$^{-1}$. The dependence on redshift is due to the evolution of the contribution to X-ray luminosity by X-ray binaries \citep[see][and references therein]{Lehmer2016}. It should be noted that the relation is expected to overestimate the X-ray luminosity of sources with SFR\,$<10$\,M$_{\odot}$\,yr$^{-1}$ \citep{Lehmer2016}. Equation \ref{Eq_SFR_X} implies that for the redshift range of SC4K LAEs ($2<z<6$) and SFRs as low as $\sim5$\,M$_{\odot}$\,yr$^{-1}$ we expect L$_{\rm X}\sim10^{41}$\,erg\,s$^{-1}$. Only SFRs of $\sim1000$\,M$_{\odot}$\,yr$^{-1}$ or higher can reach L$_{\rm X}\sim10^{42}$\,erg\,s$^{-1}$, justifying the commonly used X-ray luminosity above which AGN dominate the emission.

\section{The X-ray properties of LAEs at 2$<$z$<$6}\label{X-ray_results}

Using the method detailed in Section \ref{X-rays} we find a total of 254 (7\%) LAEs which are directly detected in the \textit{Chandra} full band ($\rm 0.5-7.0 \, KeV$) with $\rm S/N>3$ (see e.g. Figure \ref{fig:AGN}). Of these detections, 165 have S/N equal or higher than 5. The majority (89.4\%$\pm$2.3\%) of the X-ray LAEs are detected at $z<3.5$. Our detections have moderate to high X-ray luminosity ($\rm L_{0.5-10\, keV}=$10$^{43-45}$ erg s$^{-1}$) with an average luminosity of $10^{44.07 \pm 0.01}$ erg s$^{-1}$ (see Figure \ref{fig:AGN}).

Stacking in bins of Ly$\alpha$ luminosity (including X-ray detected sources) results in robust detections for the vast majority of the bins, translating into X-ray luminosities ranging from $\rm L_{0.5-10\, keV}\sim$ 10$^{42.9}$ erg s$^{-1}$ to $\sim$ 10$^{44.2}$ erg s$^{-1}$.
On the other hand, stacking by excluding X-ray sources produces no detections in general (see Figure \ref{fig:Stacks} and table \ref{table:appendix_stacks}). This result is the same if we only take out the sources from \cite{ChandraLegacy}, but some of the stacks (e.g. $z=2.9$, $z=3.7$) yield tentative detections with S/N$\sim 2$ and a X-ray luminosity of $\sim$ 10$^{42.7}$ erg s$^{-1}$. Furthermore, we note that if we only exclude the sources from \cite{ChandraLegacy}, we get significant detections for the X-ray stacks of the most luminous LAEs with luminosities L$_{\rm Ly\alpha}>10^{43.3}$\,erg\,s$^{-1}$. This is because at the highest Ly$\alpha$ luminosities there are still a significant number of X-ray sources individually detected at S/N\,$\sim3-5$ which are not in the high significance \cite{ChandraLegacy} X-ray selected catalogue.

 %%%%%%%%%%%%%%%%%%%%%%%%%%%%%%%%%%%%%%%%%%%%%%%%%%%%%
% Figure 7 - X-ray luminosity vs redshift
%%%%%%%%%%%%%%%%%%%%%%%%%%%%%%%%%%%%%%%%%%%%%%%%%%%%%
\begin{figure*}
\centering
\includegraphics[width=14.7cm]{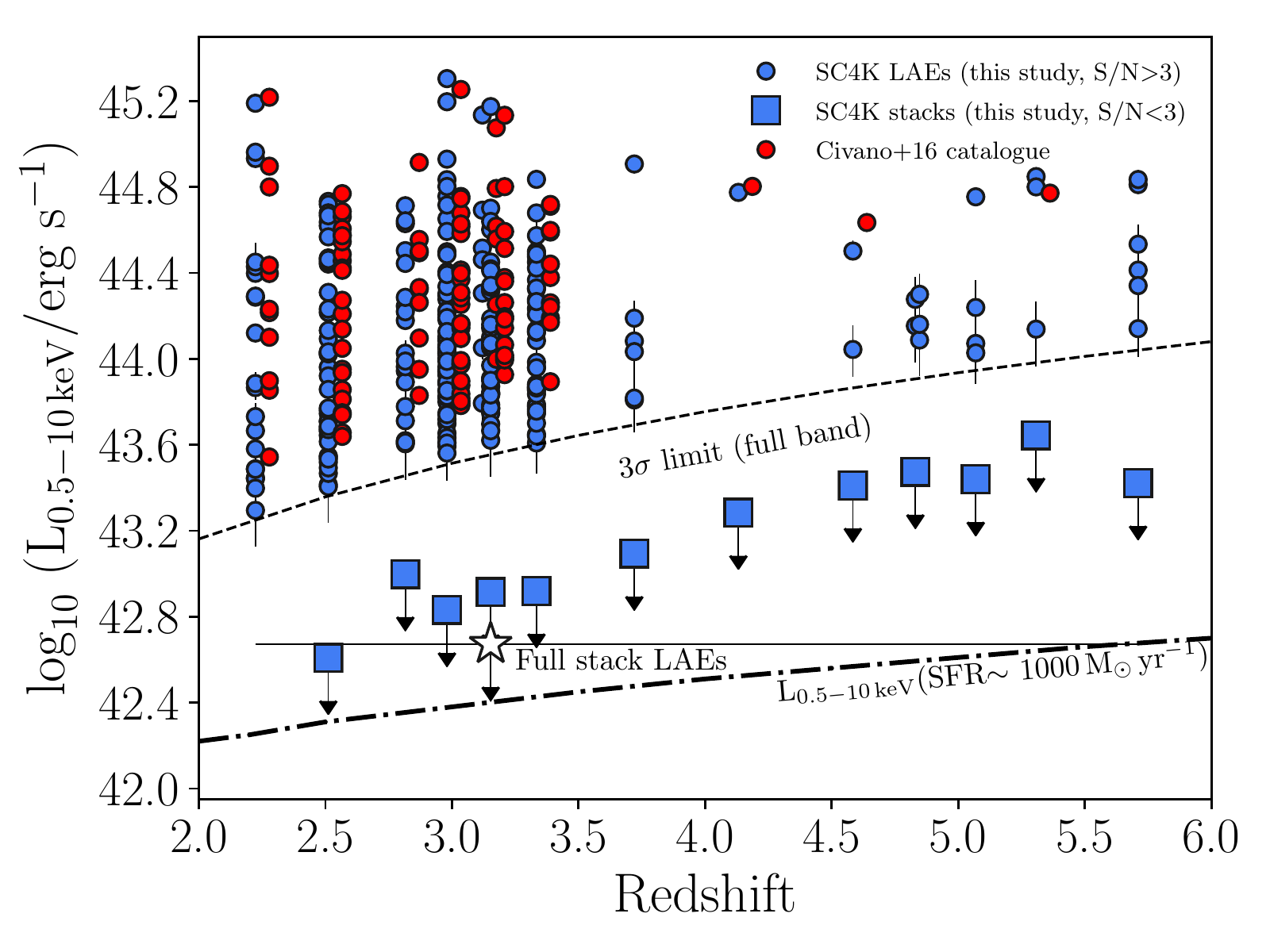}
\caption{The X-ray luminosity of all our X-ray LAEs against redshift. Stacking by excluding X-ray LAEs results in non-detections in the X-rays for all redshifts (blue squares) including the full stack (white star), which supports most non-X-ray LAEs having very low accretion rates. The black dashed line indicates the 3\,$\sigma$ luminosity detection limit used in this work. X-ray LAEs are marked by blue circles. Sources detected in the \textit{Chandra} COSMOS-Legacy catalogue are represented by red circles, with the luminosities from \citet{ChandraLegacy}. We apply a slight horizontal shift of $+0.06$ to the AGN markers from \citet{ChandraLegacy} in order to facilitate inspection. The difference in the number of sources is due to the higher signal-to-noise limit used by \citet{ChandraLegacy} of $\sim$5$\sigma$ and the different extraction methods used. We also show the luminosity limit above which AGN start to dominate the X-ray emission (dot-dashed line). This is because SFRs of $\sim1000$\,M$_{\odot}$\,yr$^{-1}$ are required \citep[][]{Lehmer2016} to achieve such X-ray luminosity. Most importantly, a SFR of $\sim5$\,M$_{\odot}$\,yr$^{-1}$ (more typical of LAEs) should lead to a X-ray luminosity of just L$_{\rm X}\sim10^{41}$\,erg\,s$^{-1}$\citep[see Section \ref{XSFR} and][]{Lehmer2016}, significantly below the stacking limits.}. 
\label{fig:XLum}
\end{figure*}

\subsection{The hardness ratio of X-ray LAEs}\label{HR_results}

Of the 254 X-ray LAEs, 143 are detected in both hard ($2.0-7.0$\,keV) and soft ($0.5-2.0$\,keV) bands individually at S/N\,$>3$. As a whole, these 143 LAEs have an average HR of $-0.1^{+0.21}_{-0.17}$. As Figure \ref{fig:HR_z_Lya} shows, approximately 48\% (69 out of 143) of our LAEs can be classed as unobscured as they have a low hardness ratio of HR\,$<-0.2$. We find no significant relation between HR and Ly$\alpha$ luminosity or redshift (see Figure \ref{fig:HR_z_Lya}), although there may be a weak trend of lower HR at the highest Ly$\alpha$ luminosities and at the highest redshifts. Our results are therefore consistent with X-ray LAEs having similar column densities/obscuration at a range of Ly$\alpha$ luminosities and across redshift.

We compare our results for X-ray detected LAEs with those based on X-ray selected sources at similar redshifts. As Figure \ref{fig:HR_z_Lya} shows, X-ray LAEs show similar hardness ratios to those reported for the global X-ray AGN population \citep{ChandraLegacy}, where the average HR is $\approx-0.11$. Interestingly, \cite{ChandraLegacy} reports that the overall population of X-ray AGN in COSMOS is best described by a double gaussian peaking at HR\,$=-0.31$ and HR\,$=0.12$. Such values could be interpreted as the result of two different X-ray AGN populations, one unobscured and one obscured, as shown by \cite{Marchesi2016IRChandra}. We find no statistically significant evidence of such distribution for the X-ray LAEs, likely due to the sample being much smaller than the full X-ray AGN sample in COSMOS, particularly towards higher redshifts.

In Figure \ref{fig:HR_Xrays} we present how the X-ray hardness ratio may depend on X-ray luminosity for LAEs. We find a significant correlation between the HR and X-ray luminosity, which implies that X-ray LAEs with higher X-ray luminosity have generally lower HR and likely lower column densities/less obscuration. This trend is very similar to what has been found by \cite{Marchesi2016} for the entire sample of X-ray AGN. Specifically, \cite{Marchesi2016} found that 80\% of the X-ray sources in COSMOS with L$\rm _X<$10$\rm ^{43}$ erg s$^{-1}$ are likely obscured AGN, while such fraction declines to $\sim$20\% at L$\rm _X>$10$\rm ^{44}$ erg s$^{-1}$. This trend has also been observed in several other studies \citep[e.g.][]{Lawrence1982,Hasinger2008,Merloni2014}, for populations with obscuration determined both optically and through X-rays \citep[but see e.g.][]{Lusso2013}, so our results seem to indicate that AGN LAEs follow the trend set by the general X-ray AGN population. Nevertheless, it is worth noting that our results may be biased towards high HR values at lower X-ray luminosities, due to requiring detections in both bands to determine the HR. As can be seen in Figure \ref{fig:HR_Xrays}, the decrease of HR with X-ray luminosity is much milder above $\sim10^{44}$\,erg\,s$^{-1}$, where in principle we are much more complete to the full range of sources regardless of their obscuration.

%%%%%%%%%%%%%%%%%%%%%%%%%%%%%%%%%%%%%%%%%%%%%%%%%
% Figure 8 - X-ray luminosity vs Lyalpha
%%%%%%%%%%%%%%%%%%%%%%%%%%%%%%%%%%%%%%%%%%%%%%%%%
\begin{figure*}
\centering
\includegraphics[width=17.9cm]{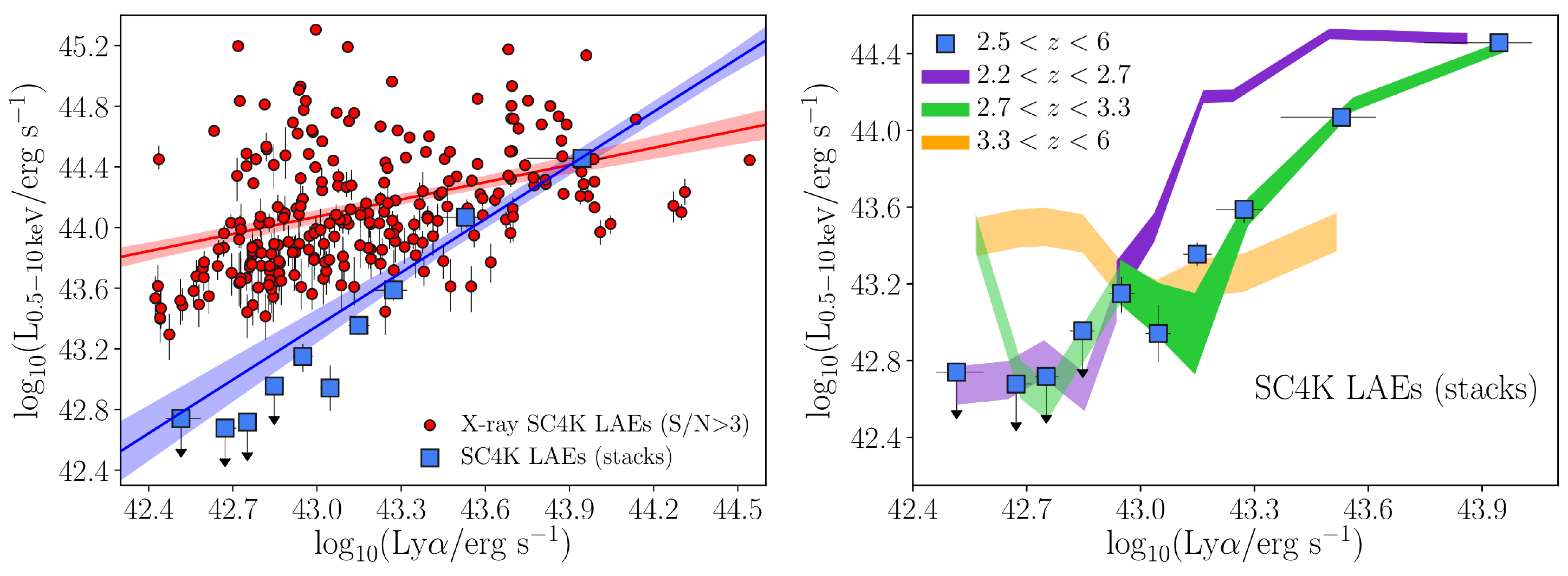}
\caption{The X-ray luminosity plotted against the Ly$\alpha$ luminosity. \textit{Left:} The red circles are the LAEs with X-ray S/N$>$3. The blue squares encompass the sources of the full SC4K sample (stacks of Ly$\rm \alpha$ luminosity bins) and the blue line represents the linear fit to the X-ray stacks as a function of Ly$\alpha$ luminosity, which results in a relation of the form $\rm log_{10}(L_X)=log_{10}(Ly\alpha) \times (1.18 \pm 0.12)-(7.3 \pm 5.3)$. The red line is a linear fit to the direct detections. We find a significant correlation, suggesting that the Ly$\alpha$ and X-ray are tracing the same physical processes. \textit{Right:} The X-ray luminosity vs the Ly$\alpha$ luminosity for LAEs at $2.2<z<2.7$, $2.7<z<3.3$ and $3.3<z<6$. The results show that LAEs at low redshift seem to have higher X-ray luminosity at a fixed Ly$\alpha$ luminosity above $10^{43}$\,erg\,s$^{-1}$.  We mark the luminosities for which we have only X-ray lower limits by increasing the transparency of the shadows (see also Table \ref{table:appendix_stacks} and the full tables available online). All Ly$\alpha$ bins at $3.3<z<6$ also provided lower limits for the X-ray luminosity.}
\label{fig:LXLy}
\end{figure*}

\subsection{X-ray luminosity of LAEs as a function of redshift}\label{Xlum_z_results}

In Figure \ref{fig:XLum} we show the X-ray luminosities of the X-ray detected LAEs as a function of redshift. The black line in Figure \ref{fig:XLum} represents the X-ray luminosity for the 3\,$\sigma$ limit in our study. The range of X-ray luminosities is relatively wide, particularly at $2<z<3$. Most X-ray LAE AGN in our sample have luminosities L$_{\rm X} \sim10^{43-45}$\,erg\,s$^{-1}$. In general, our X-ray LAEs fall within the expected luminosities for moderate to powerful AGN at similar redshifts \citep[see][and references therein]{Brandt2015}. The X-ray luminosity ranges of the AGN LAEs also correspond to high to moderate BHARs, with the highest being $\sim$4.2 M$_{\odot}$ yr$^{-1}$ and the lowest $\sim$0.04\,M$_{\odot}$ yr$^{-1}$ (see Table \ref{table:detsourc} for the full sample of our X-ray AGN candidates' properties).

We find a significant number of X-ray LAEs below $z\sim3.5$, but the number of such sources drops sharply for higher redshifts. This is partly explained by the number of LAEs at high redshift being lower than at $z<3.5$ in SC4K and the fact that the X-ray luminosity limit rises, but those alone are not the full explanation (see Section \ref{AGN_fraction_Lya}). Interestingly, the low number of X-ray LAEs is quite striking when using the high significance X-ray selected catalogue of \cite{ChandraLegacy}, where only 3 X-ray LAEs in a total of 766 are found at $z>3.5$. However, when we go down in S/N we find a significantly larger number of X-ray LAEs at high redshift, as can be seen in Figure \ref{fig:XLum}, much closer to the detection limit.

The observed low number of X-ray LAEs detected at high redshift (particularly using \citealt{ChandraLegacy}) is in agreement with the literature. \cite{Wang2004}, for example, studied $\sim 400$ LAEs at $z \sim 4.5$ (101 of which had {\it Chandra} coverage) and found no significant X-ray emission, which the authors interpreted as evidence for a lack of AGN in LAEs at that redshift. In comparison, in our study we find only 27 (0.7\% of the total SC4K sample) LAEs with X-ray emission at $\rm z>4$. This is within 1$\sigma$ of what one would expect for a distribution where we measure 0 X-ray AGN out of 101 LAEs. It is therefore likely that the low number of LAEs in \cite{Wang2004} combined with their low luminosity, was simply not enough to allow for the detection of X-ray LAEs. Fortunately, the SC4K sample of LAEs in COSMOS is finally large enough to find these rare sources even at high redshift, for the first time.

To further investigate the relative lack of X-ray detections at high redshift and any potential evolution in the population, we set up a test where we randomly picked a LAE at lower redshift ($\rm z=2.2-3.5$) and shifted it towards the highest redshifts ($\rm z=3.5-5.8$). We make sure we mimic the selection limits at higher redshift by e.g. selecting LAEs at $z\sim2-3$ with L$_{\rm Ly\alpha}>10^{43}$\,erg\,s$^{-1}$, the typical luminosity limit at higher redshift. We then find that the X-ray luminosities for the shifted sources that are detectable in {\it Chandra} would be expected to have $\rm L_X=10^{43.7-44.6}\,erg\,s^{-1}$, which is consistent with what we observe in the actual sample with our analysis. However, if the X-ray AGN fraction remained constant we would expect to have found many more X-ray LAEs at high redshift, which reveals an evolution in the population.

Results based on the general X-ray selected population at high redshift find a negative evolution of the X-ray luminosity function (XLF) where the low-luminosity end progressively lowers and flattens with redshift \citep[e.g.][]{Silverman2008, Georgakakis2015}. Given that X-ray LAEs seem to roughly reflect or follow the general X-ray selected population, it is possible that the behaviour of the XLF explains the lower number of detections at $z>3.5$. We also point out that \textit{Chandra}'s sensitivity is such that only the brighter X-ray AGN are detected even at the lowest redshifts, so it is possible we are still missing X-ray LAEs even at lower redshift (see also Section \ref{AGN_fraction_results} for further discussion).

In an effort to probe the X-ray activity of the sources which remain individually undetected in the X-rays in our analysis, we stack them. Figure \ref{fig:XLum} shows the results of X-ray stacking our sample of LAEs in bins of redshift (blue square markers) after removing individual AGN detections (X-ray S/N$>$3). We find no X-ray detection in any of the redshift binned stacks, even if we re-define the bins to encompass larger redshift intervals (see Table \ref{table:stacksourc}) or when stacking the entire sample. The X-ray upper limits we find indicate that most LAEs have no significant X-ray emission and thus LAEs are mainly SF galaxies with a small subsample of X-ray bright, AGN-powered LAEs. We note that our stacking is capable of reaching faint luminosities in the X-rays, close to $\sim10^{42}$\,erg\,s$^{-1}$ at $z=2$, the limit commonly used to separate AGN from star-forming galaxies, and corresponding to a SFR of $\sim1000$\,M$_{\odot}$\,yr$^{-1}$. We note that very low luminosity AGN can still escape detection even in our stacking analysis. We also note that rejecting only the sources from \cite{ChandraLegacy} (or those in our S/N\,$>5$ analysis) results in tentative stack detections, particularly for the stacks at $\rm z=2.9$ and $\rm z=3.7$ which reach a S/N\,$\sim2.2$. This is due to the stricter cut applied on the \textit{Chandra} catalogue, causing some of the lower luminosity LAEs that are weakly detected in the X-rays at S/N$=3-5$ to contribute to the stacks.

%%%%%%%%%%%%
% Table 1 - Stack properties
%%%%%%%%%%%%
\begin{table*}
\centering
\caption[]{The properties of the stacked LAEs in the SC4K sample as determined in this study. We divide the sample in bins of redshift and extract properties for each stack. We stack both using the full sample and excluding the X-ray and radio-detected LAEs. We also show the number of sources detected in the FIR \citep[][]{Shuowen2018} and radio. We include the median Ly$\alpha$ luminosity of the stacks, as well as the X-ray and radio mean luminosities, the SFR as determined from the Ly$\alpha$ and radio luminosities and the BHAR as determined from the X-rays. In addition, we show the BHAR/SFR ratio where the average ratio's errors are $^{+0.005}_{-0.003}$, showing the evolution of the relative black hole-to-galaxy growth. The SFRs considered for the BHAR/SFR ratio are the average between the radio and Ly$\alpha$ SFRs. When a stack leads to a S/N\,$<3$ we provide the 3$\sigma$ limit as the upper-limit for the quantities. We also include the (X-ray $+$ radio) AGN fraction, estimated using the number of sources on each bin.}

\begin{tabular}{@{}cccccccccccc@{}}
\hline
		Subsample &$\rm \log_{10}$ L& SFR& $\rm \log_{10}$ L & $\rm \log_{10}$ L&  SFR &$\dot{M}_{\rm BH}$& $\dot{M}_{\rm BH}$ & Total AGN& Radio & FIR\\
		Stacked  &Ly$\alpha$ & (Ly$\alpha$) & X-rays & radio& (radio) &(X-rays) & ------ & fraction&detected& detected \\
		(Full sample) &  [erg\,s$^{-1}$] &[M$_{\odot}$\,yr$^{-1}$]&[erg\,s$^{-1}$] &[W Hz $^{-1}$]& [M$_{\odot}$\,yr$^{-1}$]&[M$_{\odot}$\,yr$^{-1}$] & ${\rm SFR}$  &(\%)& (\#) & (\#)   \\
		\hline
		\hline
		2.2$<$z$<$2.7 & 42.6$^{+0.2}_{-0.2}$&4.2$^{+4.0}_{-2.0}$ &43.12$^{+0.05}_{-0.05}$ & 23.81$^{+0.01}_{-0.01}$& - & 0.047$^{+0.005}_{-0.005}$& 0.007  &  9.1$\pm {0.9}$ & 32 &13\\
		2.7$<$z$<$3.5 & 42.9$^{+0.2}_{-0.1}$&6.1$^{+5.5}_{-2.6}$ &43.12$^{+0.07}_{-0.07}$ & 23.22$^{+0.04}_{-0.03}$& - &  0.047$^{+0.008}_{-0.007}$& 0.005  &  9.7$\pm {0.6}$& 69 &26\\
		3.5$<$z$<$5.8 & 43.1$^{+0.2}_{-0.3}$&9.8$^{+9.9}_{-5.2}$ &$<$43.2 & $<$23.2& - & $<$0.059& $<$0.006  &  4.9$\pm {0.7}$& 15 &7\\
		\hline
		2.2$<$z$<$5.8 & 42.9$^{+0.3}_{-0.2}$&6.0$^{+7.0}_{-2.7}$ &43.06$^{+0.06}_{-0.07}$ & 23.53$\pm$0.01 & - & 0.041$^{+0.006}_{-0.006}$ &  0.005  &  8.6$\pm$0.4 & 116 & 46\\
		\hline
		(no AGN) & & &&  &    &&  \\
		\hline
		2.2$<$z$<$2.7 & 42.6$^{+0.2}_{-0.1}$&4.1$^{+3.7}_{-1.9}$ & $<$42.6& 22.45$^{+0.12}_{-0.14}$ & 9.0$^{+3.0} _{-2.5}$ & $<$0.013 & $<$0.0019 &  - & - &3\\
		2.7$<$z$<$3.5 & 42.9$^{+0.2}_{-0.1}$& $6.0^{+5.3}_{-2.5}$& $<$42.8& 22.52$^{+0.14}_{-0.14}$ & $10.6^{+4.1} _{-3.2}$ &  $<$0.021 & $<$0.0025  &  - & - &7\\
		3.5$<$z$<$5.8 & 43.1$^{+0.2}_{-0.3}$& $9.8^{+9.7}_{-5.2}$&$<$43.2 &$<$23.1 & $<$ 43.5& $<$0.060 & $<$0.0061 &  - & - &1\\
		\hline
		2.2$<$z$<$5.8 & 42.8$^{+0.3}_{-0.2}$& 5.9$^{+6.8}_{-2.7}$&$<$42.7 &22.47$^{+0.12}_{-0.13}$& 9.3 $^{+3.0} _{-2.4}$ & $<$0.017 & $<$0.0022  &  - & - &11\\
		\hline
		\label{table:stacksourc}
	\end{tabular}
\end{table*}

\subsection{X-ray luminosity vs Ly$\alpha$ luminosity}\label{BHAR_Lya_results}

To test for a relation between the X-ray and Ly$\alpha$ luminosities, we divided the sample in bins of Ly$\alpha$ luminosity and performed stacking in the X-ray full band (0.5-10\,keV). Stacking in the X-rays based on bins of Ly$\alpha$ luminosity by including the X-ray LAEs (see Figure \ref{fig:LXLy}, left panel) yields detections for all stacks except for the faintest Ly$\alpha$ luminosities ($<10^{42.9}$\,erg\,s$^{-1}$). There is a clear positive correlation between the X-ray and Ly$\alpha$ luminosities. Furthermore, the relation is present when considering the individually detected LAEs (red markers and red linear fit, Figure \ref{fig:LXLy}, left panel). From these results, it is clear that the driving force behind the X-ray-Ly$\alpha$ relation is the AGN activity. In other words, the Ly$\alpha$ emission of the X-ray direct detections is likely coming from the BH activity, and thus tracing the black hole accretion rate, while for the remainder of the sample Ly$\alpha$ likely comes from SF processes. This dichotomy of the origins of Ly$\alpha$ emission has also been identified in a recent study by \cite{Dittenber2020} on a sample of spatially resolved LAES at $\rm z<0.1$. \cite{Dittenber2020} find that, in 9 of the 12 galaxies considered, compact objects may be a major source of Ly$\alpha$ emission, with SFR processes like the activity of high mass x-ray binaries and AGN nuclear activity having possible roles in the powering of Ly$\alpha$ emission. This suggests that the duality of Ly$\alpha$ emission's origins is observed across redshift and might have consequences for the study of the epoch of reionisation.

%%%%%%%%%%%%%%%
% Figure 9 - BHAR vs Redshift
%%%%%%%%%%%%%%%
\begin{figure}
\centering
\includegraphics[width=8.4cm]{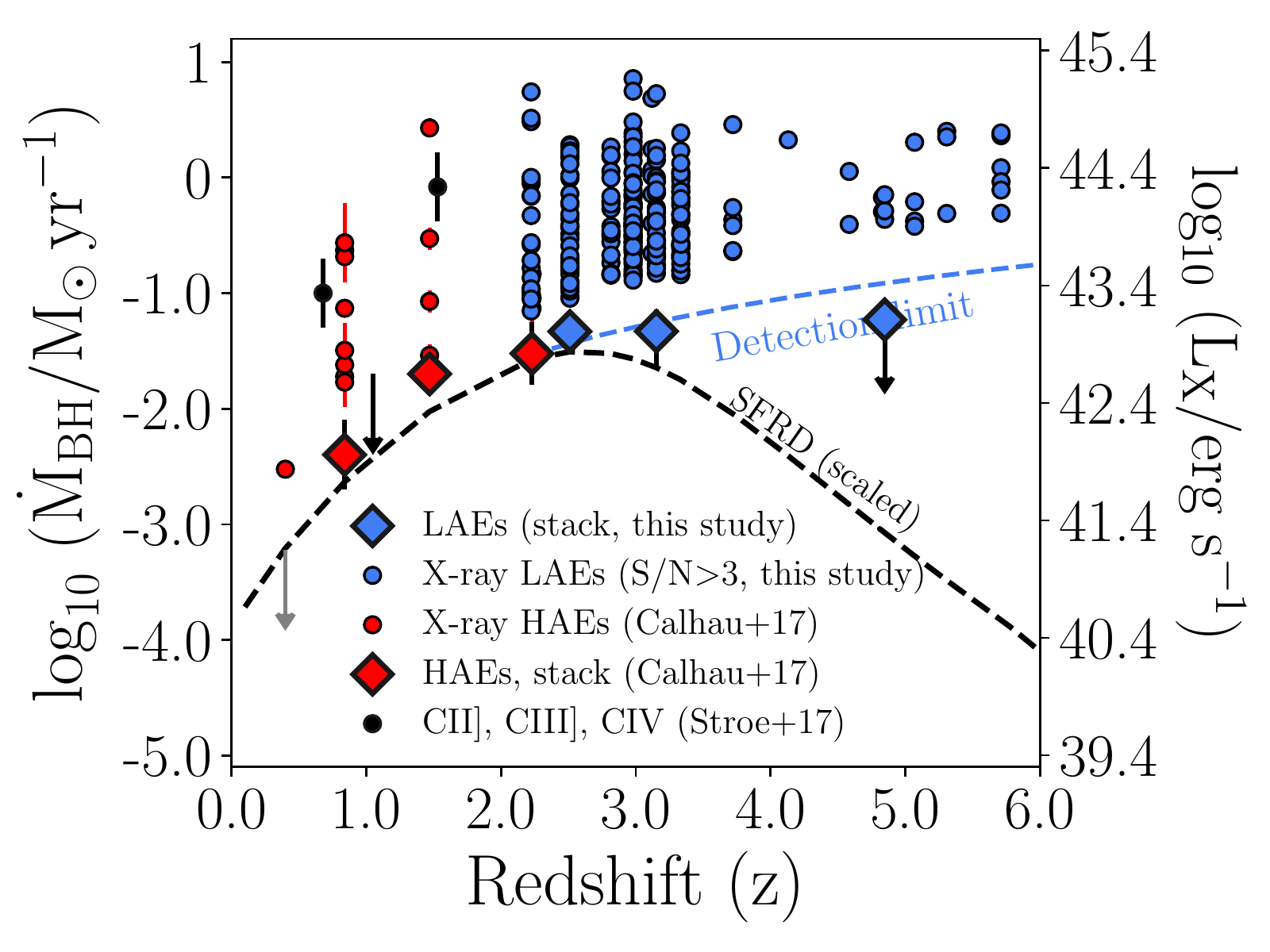}
\caption{The BHARs of LAEs across redshift. We find relatively constant BHARs for LAEs across time. The large markers represent the stacking and the smaller markers are LAEs that have been directly detected by {\it Chandra}. The dashed blue line is the detection limit of our study when demanding a 3\,$\sigma$ cut. The dashed black line shows the evolution of the SFRD derived by \citet{Khostovan2015}, scaled to coincide with the BHAR at $z\sim2.5$. Note that while the BHARs seem to follow the evolution of the SFRD up to $ z\sim2.5$, this is not clear at $z>3$, although our results suggest that there is no significant rise. We also place our results into context by comparing them with the BHARs of H$\alpha$-selected sources from \citet{Calhau2017} and to the CII$]$, CIII$]$ and CIV emitters from \citet{Stroe2017a}.}
\label{fig:redshift}
\end{figure}

We also investigate whether the X-ray-Ly$\alpha$ relation is evolving with redshift. The results are shown in  Figure \ref{fig:LXLy}, where the right panel reveals an evolution of the luminosity relation for each of the three redshift bins shown in Table \ref{table:stacksourc} ($z=2.2-2.7$, $ z=2.7-3.5$ and $ z=3.5-5.8$). Our results suggest differences in the relation that are dependent on the redshift intervals being considered. The X-ray LAEs at $2.2<z<2.7$ reveal higher X-ray luminosities for the same Ly$\alpha$ luminosity (for L$_{\rm Ly\alpha}>10^{43}$\,erg\,s$^{-1}$) when compared to the sources at $2.7<z<6$. Such result could point to an evolution in the accretion efficiency, in the typical Eddington ratios (affecting X-ray emission) and/or an evolution on the production and escape of Ly$\alpha$ photons for a given X-ray luminosity or BHAR.

\subsection{$\rm \dot{M}_{BH}$ of LAEs vs $\rm \dot{M}_{BH}$ of HAEs}\label{LAESvsHAES}

As can be seen in Figure \ref{fig:redshift}, X-ray LAEs have moderate to strong BHARs, with a median of $\rm 0.42^{+0.7}_{-0.2}\, M_{\odot}\, yr^{-1}$ and an average BHAR of 0.72$\rm ^{+0.02}_{-0.01}\, M_{\odot}\, yr^{-1}$. We also show the average stacks of our sample in three redshift bins (see also Table \ref{table:stacksourc}). Our stacking reveals relatively low BHARs of $\rm BHAR=0.047 ^{+0.005}_{-0.005}\, M_{\odot}\, yr^{-1}$ and 0.047$\rm ^{+0.008}_{-0.007}\, M_{\odot}\, yr^{-1}$, for $2.2<z<2.7$ and $2.7<z<3.5$, respectively. We find no significant detection for the $3.5<z<6$ X-ray stack (see Figure \ref{fig:redshift}).

Also shown in Figure \ref{fig:redshift} are the emission line-selected sources from \cite{Stroe2017a} and \cite{Calhau2017}. A comparison between results reveals that BHARs of X-ray LAEs are similar to the BHARs of $z=0.68$ and $z=1.53$ C{\sc{ii}]} and C{\sc{iv}} emitters (obtained by \citealt{Stroe2017a}). They are also comparable to the more powerful X-ray counterparts of HiZELS at $0.8 \leq z \leq 2.23$. Comparing the stacks reveals BHARs comparable to the BHAR for the stacks of the HAEs at redshift $\rm 1.5-2.2$. 

Figure \ref{fig:redshift} also shows the evolution of the SFRD presented by \citealt{Khostovan2015} and scaled so that the SFRD at $z\sim2.5$ coincides with the BHAR at the same redshift. Our results can be considered consistent with the BHAR following the evolution of SFRD up to $\rm z\sim3.0$, something that is mirrored by HAEs at lower redshift. Even though the current detection limits do not allow us to confirm whether such trend remains for redshifts greater than $\rm z=3.5$, the relatively strong upper limits at high redshift are consistent with a drop of the BHARs.

%%%%%%%%%%%%%%%%%%%%%%%%%%%%%%%%%%%%%%%%%%%%%%%%%%%%%
% Figure 10 - Radio Stacks
%%%%%%%%%%%%%%%%%%%%%%%%%%%%%%%%%%%%%%%%%%%%%%%%%%%%%
\begin{figure}
\centering
\includegraphics[width=8.4cm]{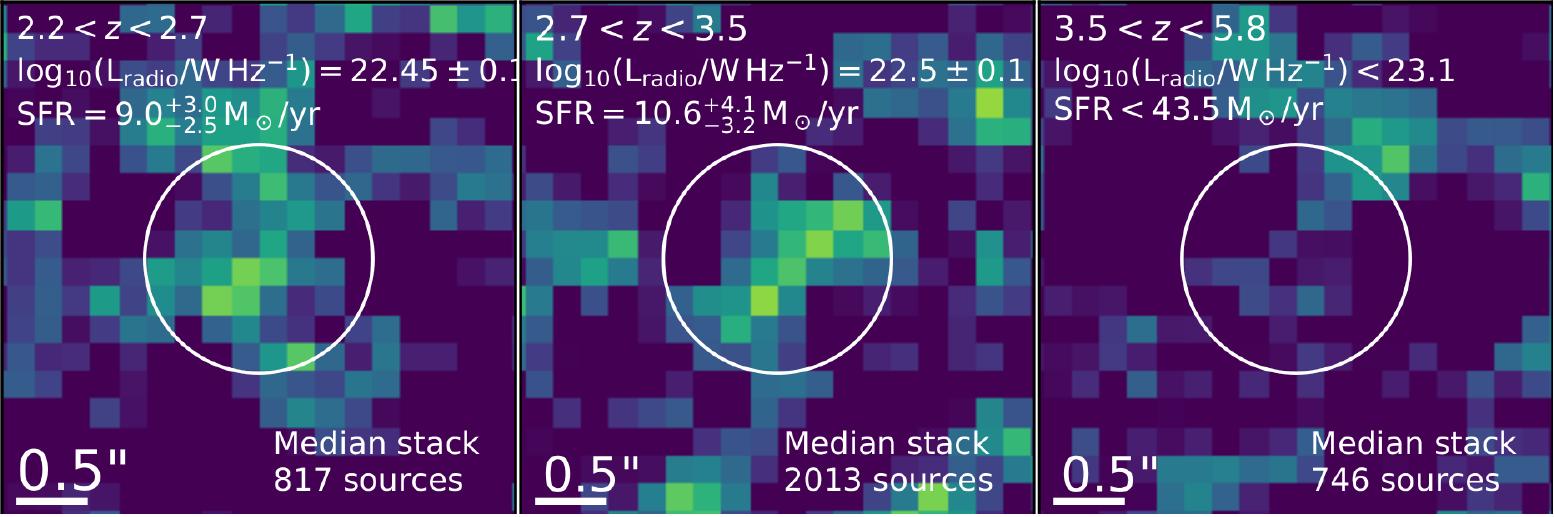}
\caption{The results of median stacking the SC4K LAEs not directly detected in the radio. We detect weak radio emission from LAEs for the stacks at $z=2.2-2.7$ ({\it left}) and $z=2.7-3.5$ ({\it middle}) with radio luminosities of L$_{\rm radio}=10^{22.5\pm0.1}$\,W\,Hz$^{-1}$, corresponding to SFR\,$= 9.0^{+3.0}_{-2.5}$\,M$_{\odot}$\,yr$^{-1}$ and L$_{\rm radio}=10^{22.5\pm0.1}$\,W\,Hz$^{-1}$, corresponding to SFR\,$= 10.6^{+4.1}_{-3.2}$\,M$_{\odot}$\,yr$^{-1}$, respectively. For $z>3.5$ we are able to provide a 3\,$\sigma$ upper limit of SFR\,$<44$\,M$_{\odot}$\,yr$^{-1}$.}
\label{fig:Stacks_radio}
\end{figure}

\section{The radio properties of LAEs at 2$<$z$<$6}\label{radio_results}

Using the method detailed in Section \ref{radio}, we find a total of 116 LAEs (S/N$>$3) in either the 1.4\,GHz or 3\,GHz radio bands. Out of the total 116 radio sources, most (88) are detected in the 3\,GHz data, with 28 being detected exclusively in the 1.4\,GHz VLA data and 25 in both. Out of all radio sources, 56 are also detected in the X-rays by {\it  Chandra}.

We obtain a very significant detection (S/N\,$\sim50$) when stacking the entire sample of LAEs in the radio, revealing a radio luminosity of L$_{\rm radio}=10^{23.53\pm0.01}$\,W\,Hz$^{-1}$. We also find detections in the radio stacks of LAEs when we split the sample  in redshift, with stacks of LAEs at $z=2.2-2.7$ and $z=2.7-3.5$ yielding particularly high S/N radio detections. Stacks of LAEs obtained as a function of Ly$\alpha$ luminosity also reveal clear detections, but these include the direct radio detections, almost certainly powered by AGN activity, which we find dominate the stacks.

Removing the LAEs directly detected in the radio from the sample leads to a much lower radio signal, and to a weak ($\rm S/N=3.9$) radio detection for the entire sample, with L$_{\rm radio}=10^{22.47\pm0.1}$\,W\,Hz$^{-1}$. We also detect weak radio emission when removing radio detections for $2.2<z<2.7$ and $2.7<z<3.5$ (see Figure \ref{fig:Stacks_radio}), but not at the highest redshifts (see Table \ref{table:stacksourc}).

\subsection{Radio spectral index and Ly$\alpha$ luminosity}\label{alpha_results}

\begin{figure}
\centering
	\includegraphics[width=8.2cm]{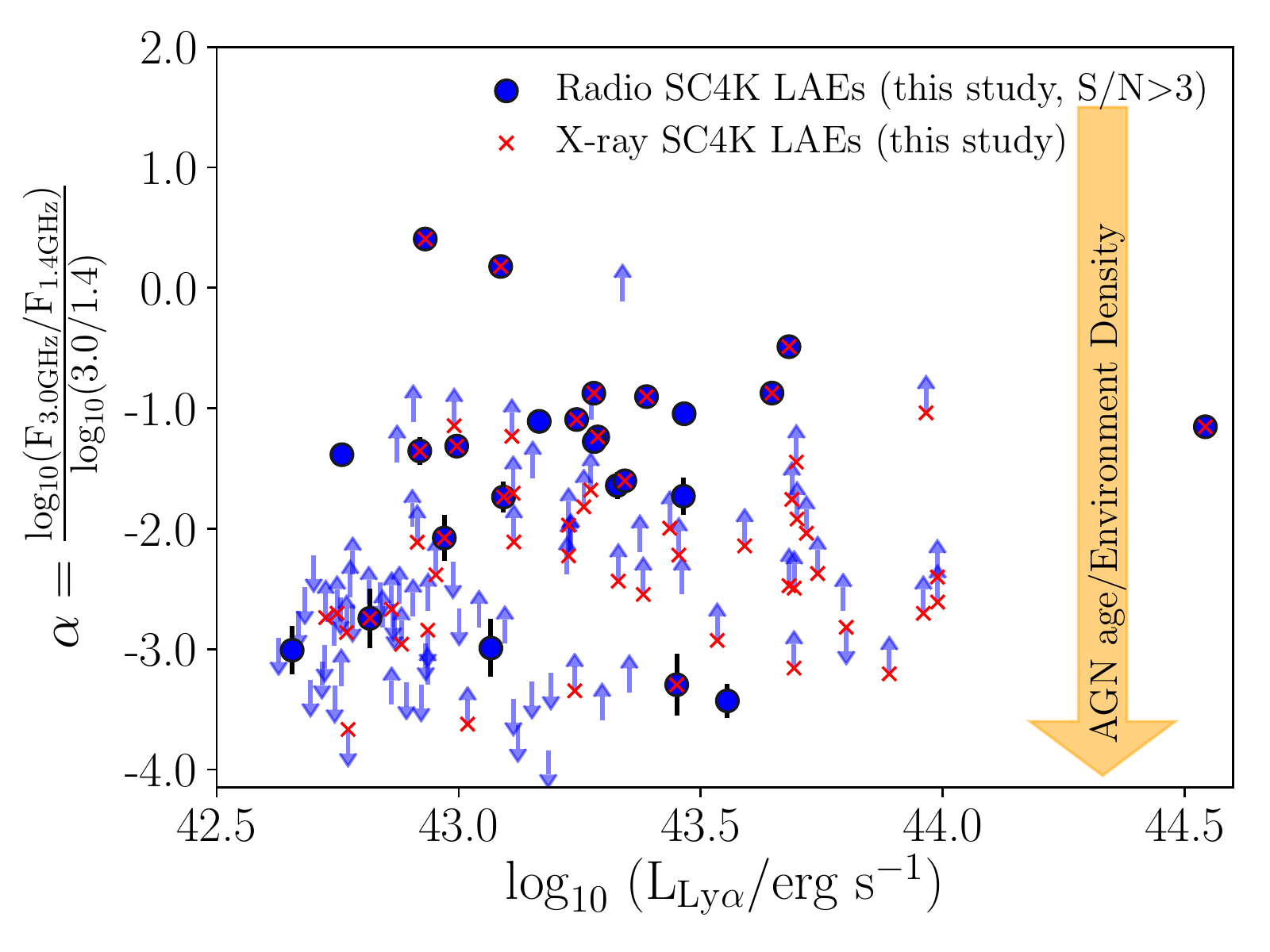}
	\caption{The radio spectral index, estimated between 1.4\,GHz and 3.0\,GHz, against the Ly$\alpha$ luminosity for LAEs. Our results show that there is no significant relation between the radio spectral index and Ly$\alpha$ luminosity, consistent with radio properties being uncorrelated with Ly$\alpha$ properties for LAEs.}
	\label{fig:radioalpha}
\end{figure}

We follow Section \ref{radio_alpha} and estimate the radio spectral index for the LAEs detected in both the 1.4 and 3.0\,GHz bands. We find that the average spectral index is $-1.3^{+0.4}_{-1.5}$. We find some unusually steep spectral indices, especially for the sources for which we can only provide limits (see Figure \ref{fig:radioalpha}), but many of these values are affected by large errors. We note, nonetheless, that \cite{Smolcic2017} constrain their spectral indices to a minimum of $-2.5$, when estimating $\alpha$. This is because standard synchrotron radiation does not result in spectral indices lower than $-2.5$, unless it is an exotic source \citep[see][]{Rees1967, Krishna2014}. We do not apply any constraints to our spectral indices but it should be noted that values lower than $-2.5$ are unlikely to be physically meaningful and that applying a cut of $\rm S/N>5$ to our data results in these low spectral indices disappearing for all but two sources.

As Figure \ref{fig:radioalpha} shows, we do not find a statistically significant relation between the radio spectral index ($\alpha$) and Ly$\alpha$ luminosity, suggesting little to no relation between them for the range of Ly$\alpha$ luminosities probed by SC4K. However, we find that the radio spectral indices of the radio LAEs are steeper than the mean spectral index for the overall COSMOS survey \citep[$\rm \overline{\alpha} = -0.73$, see][]{Smolcic2017}. Our results imply that radio LAEs are not representative of the overall radio-selected population. The steeper (more negative) average $\alpha$ for radio LAEs is still consistent with a sub-sample of AGN sources and potentially some more extreme star-forming galaxies in literature \citep[e.g.][]{Delhaize2017}.

Spectral indices can be used in several different ways, such as a probe for the origins of the radio emission itself \citep[e.g. thermal emission, synchrotron radiation;][]{KleinEmerson1981}. Thermal emission from H{\sc ii} regions would result in a spectral index of $\alpha=-0.1$ to $\alpha=2.0$, with steeper indices being a characteristic of synchrotron radiation. Our results are therefore consistent with radio emission from synchrotron processes, such as those found in radio AGN.  Spectral indices can also be used as a measure of the age or density of the environment surrounding the source. This is because radio galaxies with steeper spectral index are generally located at the centre of rich clusters of galaxies \citep[see e.g.][]{Athreya1998, Klamer2006}. As radio LAEs have steeper $\alpha$ than the general radio-selected population, our results may suggest that they are good tracers of high density regions at high redshift \citep[i.e. protoclusters; ][]{Franck&McGaugh2016}. This is consistent with several results in the literature. For example, \cite{Venemans2007} and \cite{Yamada2012} found bright LAEs around dense regions of the Universe for $2<z<5.2$ \citep[see also][]{Overzier2016, Kubo2013}. Furthermore, more recently, \cite{Khostovan2019} conducted a detailed clustering analysis of faint to luminous LAEs, including the SC4K LAEs, to find that luminous LAEs reside in the most massive dark matter haloes at high redshift, and are therefore consistent with being progenitors of some of the most massive clusters found today \citep[see also][]{Matsuda2004}.
%%%%%%%%%%%%%%%%%%%%%%%%%%%%%%%%%%%%%%%%%%%%%%%%%
% Figure 11 - radio Spectral Index
%%%%%%%%%%%%%%%%%%%%%%%%%%%%%%%%%%%%%%%%%%%%%%%%%

\subsection{Radio luminosity of LAEs as a function of redshift}\label{radio_z_results}

%%%%%%%%%%%%%%%%%%%%%%%%%%%%%%%%%%%%%%%%%%%%%%%%%
% Figure 12 - radio luminosity vs z
%%%%%%%%%%%%%%%%%%%%%%%%%%%%%%%%%%%%%%%%%%%%%%%%%
\begin{figure}
	\centering
	\includegraphics[width=8.4cm]{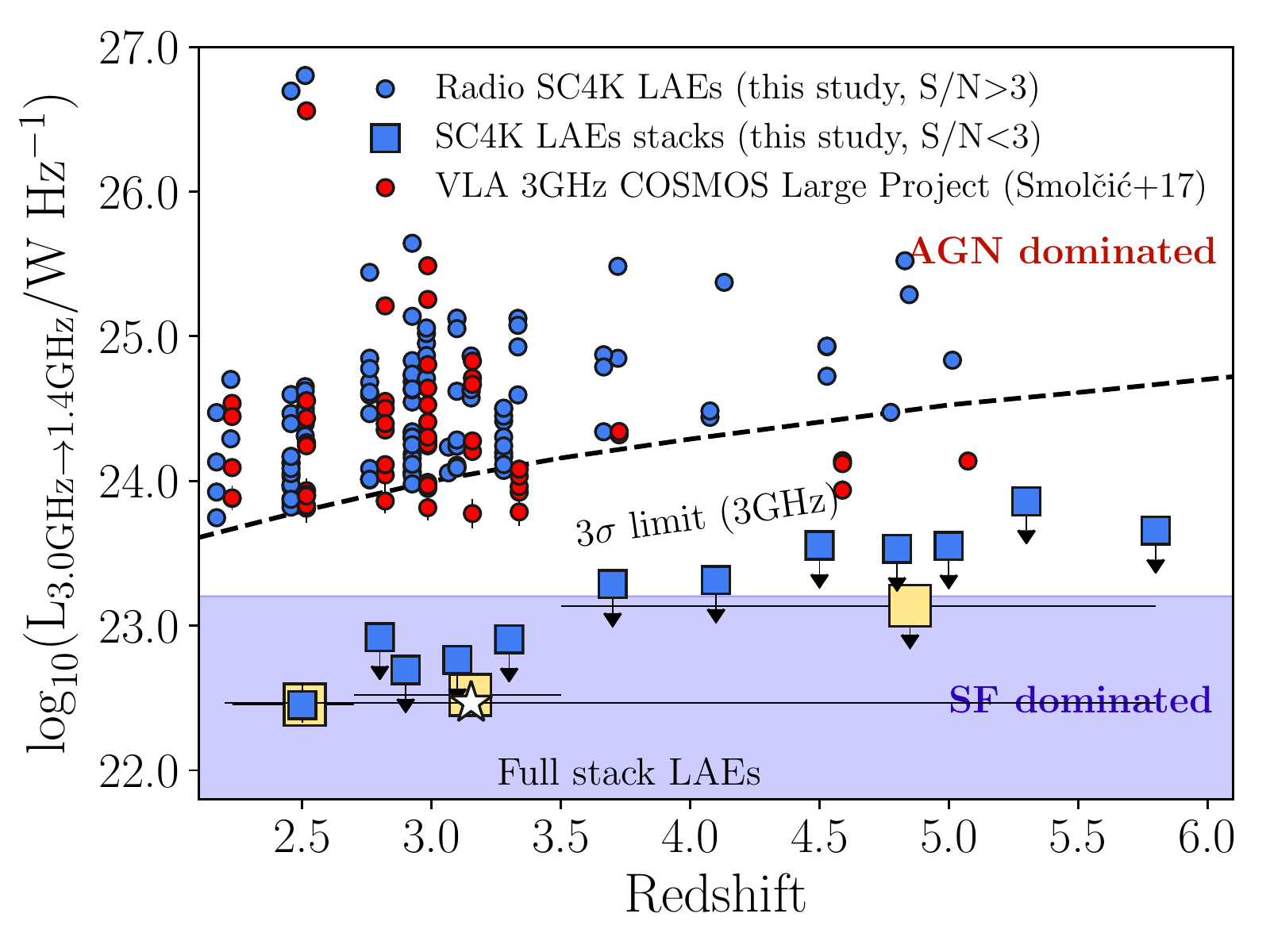}
	\caption{The radio luminosity of LAEs across redshift. The blue circles represent the direct detections found following the method presented in Section  \ref{radio}, while the red circles show our results using the VLA 3\,GHz COSMOS catalogue \citep[][]{Smolcic2017}, shifted by $+0.05$ in redshift. We also show the results of stacking in bins of redshift (square markers). We find detections for the stacks of $2.2<z<2.7$ and $2.7<z<3.5$. We also detect radio emission when stacking the full sample while excluding radio LAEs (white star). All three detected stacks are situated well in the radio luminosity range where star formation is expected to take over radio emission \citep[e.g.][]{Meurs1984}.}
	\label{fig:radiolum}
\end{figure}

The LAEs detected directly in the radio present an average L$_{\rm radio}=10^{24.94\pm0.02}$\,W\,Hz$^{-1}$ across the full redshift range, as can be seen in Figure \ref{fig:radiolum}. Such high radio luminosities are well into the AGN dominated region of the radio luminosity range \citep[L$_{\rm radio}>10^{23.2}$\,W\,Hz$^{-1}$; e.g.][see Figure \ref{fig:radiolum}]{Meurs1984,Sadler2002}. Even at $z\sim2$ the radio data is not deep enough to individually detect sources which would be in the clear SF regime. At higher redshifts the higher luminosity limit leads to a stronger bias towards detecting the highest radio luminosity LAEs only (see Figure \ref{fig:radiolum}). This high luminosity limit at high redshift, combined with a lower number of LAEs, might be able to explain the relatively low number of radio LAEs at higher redshift, although it is worth noting that we do not find a single radio LAE at a redshift beyond $z\sim5$.

The radio detection limit biases detections towards radio AGN. However, by stacking, and particularly by stacking in the radio and excluding radio LAEs, we can investigate the typical radio luminosities of the remaining population. Our results are shown in Figure \ref{fig:radiolum}. For our stacking analysis we use the 3\,GHz band, due to the higher resolution and depth of the survey. We find weak radio detections (S/N\,$\sim3-4$) in the stacks when we use the full sample of non radio LAEs and at lower redshift, placing the majority of the sources well within the SF dominated region of radio luminosities, with L$_{\rm radio}\approx10^{22.4-22.5}$\,W\,Hz$^{-1}$. Splitting the sample in further redshift slices leads to upper limits which are fully consistent with our weak detections, only achievable due to the combination of the radio data depth and the large number of LAEs in SC4K. Our results imply that some LAEs have high radio luminosities, allowing them to be directly detected in the radio, but that the majority of the LAE population is made of star-forming galaxies with very weak radio luminosities, only detectable with very deep radio stacks.

\subsection{Radio luminosity vs Ly$\alpha$ luminosity}\label{radio_Lya_results}

%%%%%%%%%%%%%%%%%%%%%%%%%%%%%%%%%%%%%%%%%%%%%%%%%
% Figure 13 - radio luminosity vs Lyalpha
%%%%%%%%%%%%%%%%%%%%%%%%%%%%%%%%%%%%%%%%%%%%%%%%%
\begin{figure}
	\centering
	\includegraphics[width=8.4cm]{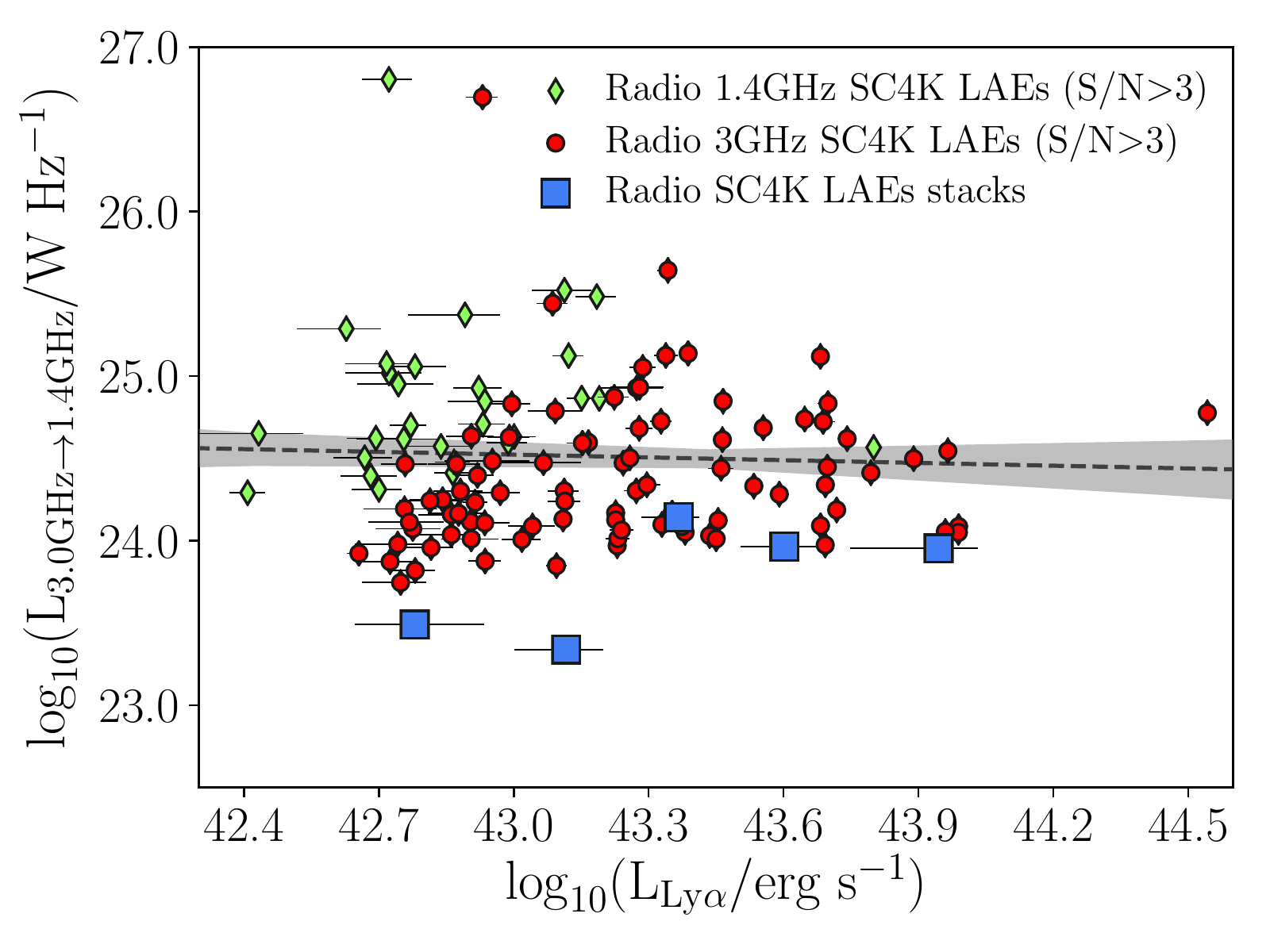}
	\caption{The radio luminosity of LAEs versus their Ly$\alpha$ luminosity. We find no statistically significant correlation between the two quantities, suggesting that radio and Ly$\alpha$ are tracing processes with different physical origins or timescales. The red markers are the LAEs detected directly in the 3.0\,GHz band, while the green diamonds represent the sources that are only detected at 1.4\,GHz. The blue squares represent stacking made in bins of Ly$\alpha$ luminosity using all LAEs. The black line represent a linear fit to direct detections with 1$\sigma$ uncertainties.}
	\label{fig:radiolya}
\end{figure}

In Section \ref{BHAR_Lya_results} we found strong relations between X-ray and Ly$\alpha$ luminosities, implying a clear link between them and BHARs. Here we investigate if there is a similar relation between radio and Ly$\alpha$ luminosities. The results are presented in Figure \ref{fig:radiolya}. Our results show a flat relation between radio and Ly$\alpha$ luminosities for radio detected sources (see Figure \ref{fig:radiolya}). Stacking (including radio LAEs) shows a potential weak relation but this result is consistent with no relation within 2$\sigma$. 

The absence of a relation between the radio and Ly$\alpha$ luminosities suggests radio emission and Ly$\alpha$ emission may be unrelated or out of sync, unlike X-ray and Ly$\alpha$. It is possible that the radio is simply tracing different AGN-related processes than the ones Ly$\alpha$ and X-rays trace. Differences could arise if the origin of radio emission happens on different physical scales (e.g. jets or away from the X-ray emitting region), also implying different timescales between the accretion of matter and the emission, but also because radio emission can be much more long-lived. Significant variability in AGN LAEs could potentially explain why radio luminosities for LAEs are uncorrelated with the likely BHAR-driven Ly$\alpha$ emission.

\section{Is there a BH-galaxy co-evolution in LAEs? AGN fractions, SFRs and BHAR/SFR}\label{AGN_fraction_results}

In total, out of 3700 LAEs, 314 are classified as AGN due to their detection in the X-ray full band or one of the radio bands. Of the 314 AGN LAEs, 254 are detected in the X-rays. We also identify 116 galaxies with detectable radio emission. This results in a total LAE AGN fraction of $8.5\% \pm 0.4\%$. We stress that this is a lower limit as there may be AGN in our sample that are too faint to be detected \citep[e.g.][]{Sobral2018}.

 %%%%%%%%%%%%%%%%%%%%%%%%%%%%%%%%%%%%%%%%%%%%%%%%%%%%%
% Figure 14 - AGN fraction vs Lya
%%%%%%%%%%%%%%%%%%%%%%%%%%%%%%%%%%%%%%%%%%%%%%%%%%%%%
\begin{figure*}
\centering
\includegraphics[width=17.7cm]{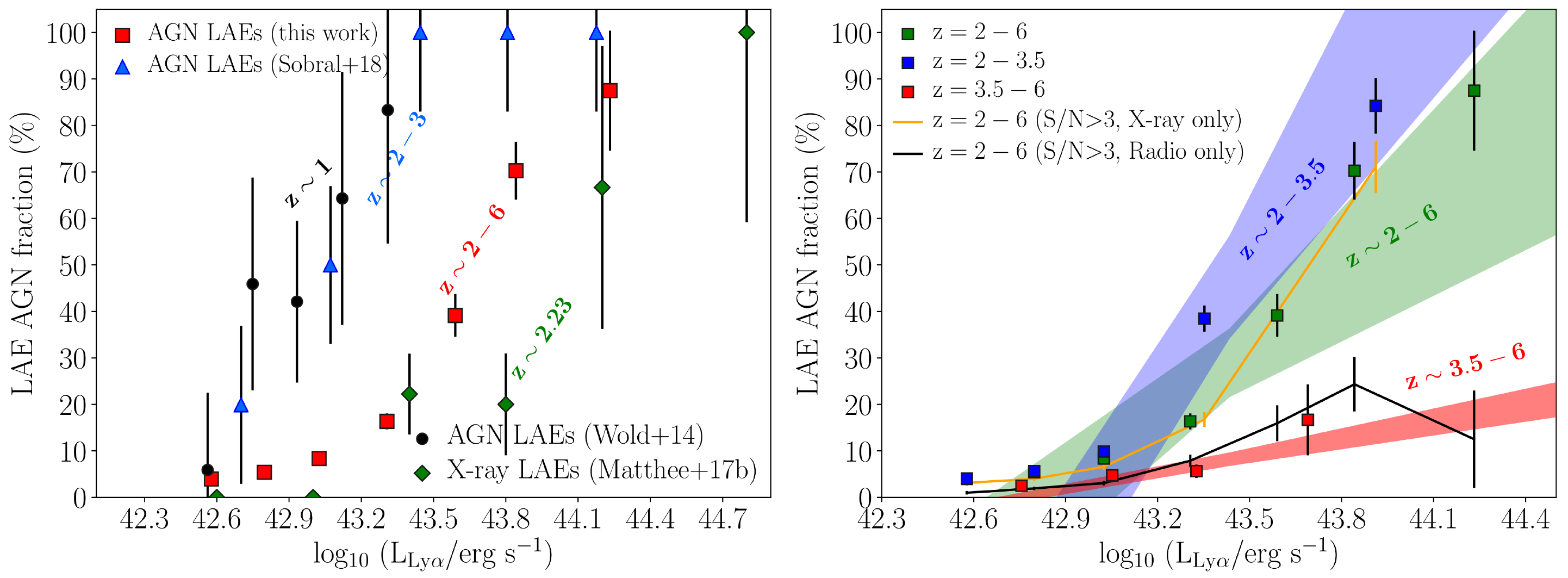}
\caption{{\it Left}: The detected LAE AGN fraction rises steeply with increasing Ly$\alpha$ luminosity for our full sample, revealing that the most luminous LAEs are almost all AGN. Our results are in agreement with those found in the literature \citep[e.g.][]{Wold2014, Matthee2017, SC4Kpaper}. {\it Right}: The evolution of the AGN fraction for the entire sample (green) and for LAEs at $\rm z=2-3.5$ (blue) and $\rm z=3.5-6$ (red). We find a significant redshift evolution of the AGN fraction as a function of Ly$\alpha$ luminosity. The fraction growth is much steeper at lower redshifts than at $3.5<z<6$. The shaded regions represent the 16th-86th percentiles of linear fits obtained when the Ly$\alpha$ luminosity bins and bin widths are varied randomly. We also show that the rise is dominated by X-ray LAEs, while the radio AGN fraction still rises with Ly$\alpha$ luminosity, but at a much shallower rate. We caution that our AGN fractions should be interpreted as lower limits as there could be undetected AGN in both radio and X-ray bands, as shown in \citet{Sobral2018}.}
	\label{fig:Histfractions}
\end{figure*}

\subsection{AGN fraction and its redshift evolution}\label{AGN_fraction_z}

The X-ray AGN fraction for LAEs is found to be $7.3\%\pm0.8\%$ at $z \sim 2.2-2.7$ and $7.9\% \pm 0.5\%$ at $z \sim 2.7-3.5$. At z$>$3.5, the X-ray AGN fraction of LAEs drops to $3.5\% \pm 0.6\%$. The decline at higher redshift is found regardless of the signal-to-noise cut employed or whether we use \cite{ChandraLegacy}'s catalogue (see Table \ref{table:AGNfractionsdifference} for full details). 

 %%%%%%%%%%%%%%%%%%%%%%%%%%%%%%%%%%%%%%%%%%%%%%%%%%%%%
% Table 2 - AGN fractions
%%%%%%%%%%%%%%%%%%%%%%%%%%%%%%%%%%%%%%%%%%%%%%%%%%%%%
\begin{table}
\centering
\caption[]{The evolution of the LAE X-ray and radio detected AGN fractions with redshift. We present the results obtained by selecting sources with a S/N\,$>3$, S/N\,$>5$ and also by using only the sources detected in \citet{ChandraLegacy} and \citet{Smolcic2017} catalogues as AGN LAEs. We find a general drop in the AGN fraction at the highest redshifts. This drop is particularly steep towards $z\sim3.5-6$ when using the high significance detection catalogues of \citet{ChandraLegacy} and \citet{Smolcic2017} and our own analysis with S/N\,$>5$. Radio AGN fractions show a flatter evolution overall when compared to X-ray LAEs. Our AGN fractions should be interpreted as lower limits as there could be undetected AGN in both radio and X-ray bands.} 
	\begin{tabular}{@{}lcccc@{}}
		%\hline
		\hline
		\hline
		X-ray LAEs &AGN fraction (\%)&&\\
		\hline
		\hline
		Sample & 2$<$z$<$2.7 &2.7$<$z$<$3.5 &3.5$<$z$<$6 \\
		\hline
		This Work (3$\sigma$) & 7.3$\pm {0.8}$ & 7.9$\pm {0.5}$ & 3.5$\pm {0.6}$\\
		This Work (5$\sigma$) & 5.2$\pm {0.7}$ & 5.1$\pm {0.5}$ & 0.8$\pm {0.3}$\\
		Civano et al.(2016) & 3.9$\pm {0.6}$ & 3.5$\pm {0.4}$ & 0.4$\pm {0.2}$\\
		\hline
		\hline
		Radio LAEs &AGN fraction (\%)&&\\
		\hline
		Sample & 2$<$z$<$2.7 &2.7$<$z$<$3.5 &3.5$<$z$<$6 \\
		\hline
		This Work (3$\sigma$) & 3.8$\pm {0.6}$ & 3.3$\pm {0.4}$ & 1.9$\pm {0.5}$\\
		This Work (5$\sigma$) & 1.3$\pm {0.4}$ & 1.2$\pm {0.2}$ & $0.8\pm {0.3}$\\
		Smol\u{c}i\'{c} et al.(2017) & 1.9$\pm {0.5}$ & 1.7$\pm {0.3}$ & 0.8$\pm {0.3}$\\
		\hline
		\hline
		Radio+X-ray LAEs &AGN fraction (\%)&&\\
		\hline		
		Sample & 2$<$z$<$2.7 &2.7$<$z$<$3.5 &3.5$<$z$<$6 \\
		\hline
		This Work (3$\sigma$) & 9.1$\pm {0.9}$ & 9.6$\pm {0.6}$ & 4.9$\pm {0.7}$\\
		This Work (5$\sigma$) & 5.5$\pm {0.7}$ & 5.6$\pm {0.5}$& 2.5$\pm {0.5}$\\
		\hline
		
		\label{table:AGNfractionsdifference}
	\end{tabular}
\end{table}

The total radio AGN fraction of LAEs is $3.1\% \pm 0.3\%$. We find that the radio AGN fraction remains relatively constant at $\sim 3.4\%$ at $z \sim 2.2-3.5$, before falling towards $2.0\%\pm 0.5\%$ at $z \sim 3.5-5.8$. We note that the uncertainties in the AGN fractions allow for the fractions to remain constant or even rise slightly between $z=2.2$ and $z=5.8$ in all cases (3$\sigma$ cut, 5$\sigma$ cut and \citealt{Smolcic2017}'s catalogue, see Table \ref{table:AGNfractionsdifference}). The radio AGN fraction among LAEs shows a much flatter evolution with redshift than the X-ray AGN fraction.

Overall, the AGN fraction of LAEs stays relatively constant up to $z\sim3.5$, from $9.1\% \pm 0.9\%$ at $z \sim 2.2-2.7$ to $9.6\% \pm 0.6\%$ at $z \sim 2.7-3.5$ before dropping by a factor of almost 2 to $4.9\% \pm 0.7\%$ at $z \sim 3.5-5.8$ (see Tables \ref{table:stacksourc} and \ref{table:AGNfractionsdifference}). Using a higher S/N cut (or the \citet{ChandraLegacy} and \citet{Smolcic2017} catalogues) leads to an even sharper decline of the AGN fraction at the highest redshifts, but the qualitative result is the same.

\subsection{The AGN fraction dependence on Ly$\alpha$ luminosity}\label{AGN_fraction_Lya}

\subsubsection{Global AGN fraction: global rise with Ly$\alpha$}

Figure \ref{fig:Histfractions} presents how the full LAE AGN fraction (radio + X-rays) varies as a function of increasing Ly$\alpha$ luminosity for SC4K LAEs. The global AGN fraction clearly rises with L$_{\rm Ly\alpha}$, to the point where the most luminous LAEs are almost, if not all, AGN. In practice, the AGN fraction rises from $\sim0-5$\% to $\sim80-100$\% from L$_{\rm Ly\alpha}\sim10^{42.6}$\,erg\,s$^{-1}$ to L$_{\rm Ly\alpha}\sim10^{44.4}$\,erg\,s$^{-1}$ at $z\sim2-6$.

The left panel of Figure \ref{fig:Histfractions} compares our results with other recent studies. \cite{Matthee2017} conducted a similar analysis to ours, albeit with shallower X-ray and Ly$\alpha$ data. They find a very similar rise in the (X-ray) AGN fraction as a function of Ly$\alpha$ luminosity at $z\sim2.2$.  \cite{Matthee2017} also notes that the increase in the fraction corresponds to the luminosity at which the number densities start to deviate from the Schechter function \citep[see also][]{Konno2016,CALYMHA2017}, which is fully captured and discussed at multiple redshifts by \cite{SC4Kpaper}. \cite{Wold2014, Wold2017} used spectroscopy to classify LAEs at $z\sim1$, finding a similar relation which is offset to much higher AGN fractions for a fixed Ly$\alpha$ luminosity. This could be interpreted as a redshift evolution, but as shown by \cite{Sobral2018} that is likely not the explanation. \cite{Sobral2018} followed-up spectroscopically (using rest-frame UV lines and photo-ionisation modelling) a sample of the most luminous $z\sim2-3$ LAEs to find a very similar relation (including the normalisation) to \cite{Wold2014, Wold2017}, consistent with no redshift evolution from $z\sim1$ to $z\sim2-3$. Instead, \cite{Sobral2018} show how X-ray data only allow to estimate lower limits for the AGN fraction at high redshift. Their results show that essentially all LAEs with  L$_{\rm Ly\alpha}>$10$^{43.3}$ erg s$^{-1}$ are AGN. This rise is extremely fast as can be seen in Figure \ref{fig:Histfractions}. 

\subsubsection{X-ray and radio AGN fractions as a function of Ly$\alpha$}

We find that the X-ray AGN fraction on its own rises steeply with Ly$\alpha$ luminosity, from $3.6\% \pm 0.7\%$ at $\rm L_{Ly\alpha}\sim10^{42.7} \, erg \, s^{-1}$ to $70\% \pm 5\%$ at $\rm L_{Ly\alpha}\sim10^{43.7} \, erg \, s^{-1}$ (see the right panel of Figure \ref{fig:Histfractions}). At the highest Ly$\alpha$ luminosities, most LAEs become detected by {\it Chandra}, showing that AGN detected in the X-rays dominate the sample at high Ly$\alpha$ luminosities. These results are also qualitatively observed if we restrict the sample to the S/N$>5$ X-ray detections or use \cite{ChandraLegacy}'s catalogued sources.

The radio AGN fraction of LAEs shows a flatter rise with Ly$\alpha$ luminosity (see Figure \ref{fig:Histfractions}), reflecting the different L$\rm _{X}$-L$\rm _{Ly\alpha}$ and L$\rm _{radio}$-L$\rm _{Ly\alpha}$ relations we find for LAEs (see Sections \ref{BHAR_Lya_results} and \ref{radio_Lya_results}). The radio AGN fraction of LAEs remains relatively low for most of Ly$\alpha$ luminosity bins, only growing to 25-30\% for the highest Ly$\alpha$ luminosities. The results support a radio AGN fraction that rises with Ly$\alpha$ luminosity but with a shallower slope and a much lower normalisation. Interestingly, we also find that the rise of the radio AGN population for the highest Ly$\alpha$ luminosities is driven by the inclusion of radio sources which are also X-ray sources. Restricting the AGN fraction to pure radio sources which remain undetected in the X-rays results in an even lower radio AGN fraction with the highest values only reaching $\sim$5\% even at the highest Ly$\alpha$ luminosities.

\subsubsection{The rise of the LAE AGN fraction with Ly$\alpha$ evolves with redshift}

\cite{Sobral2018} discuss the possibility of the LAE AGN fraction being much lower at fixed observed Ly$\alpha$ luminosity towards higher redshifts \citep[particularly based on results from][]{Matthee2017c,Matthee2017b} and the physical implications/interpretations. We can investigate this possibility for the first time by splitting our sample into a higher and lower redshift sub-sample. Figure \ref{fig:Histfractions} (right panel) shows our results. We find that the LAE AGN fraction increases with Ly$\alpha$ luminosity at both $z\sim2-3.5$ and $z\sim3.5-6$, but shows significant evolution as suggested by \cite{Sobral2018}. The AGN fraction is higher and seems to rise more steeply with increasing Ly$\alpha$ luminosity at $z\sim2-3.5$ than at $z\sim3.5-6$. While at $z\sim2-3.5$ virtually all LAEs with Ly$\alpha$ luminosities in excess of $10^{44}$\,erg\,s$^{-1}$ are (X-ray or radio) AGN, by $z\sim3.5-6$ the measured AGN fraction is only $\sim10$\%. While it is harder to detect AGN with X-rays and radio at higher redshift, the strong redshift evolution is much stronger than expected simply based on a detection bias. The decrease in the AGN fraction with redshift, for a fixed Ly$\alpha$ luminosity, may be due to the fact that the BH accretion rate density of X-ray AGN drops significantly for $z>3$ \citep[e.g. ][]{Vito2016} or due to the high fraction of obscured AGN population at high redshifts \citep[$>50\%$ at $z>3$ and increasing with redshift, see e.g. ][]{Vito2014}.

\cite{Sobral2018} argue that the relation between the AGN fraction of LAEs and their observed luminosity at $z\sim2-3$ (and the steepness of the relation) is caused by star-forming galaxies having a maximum observable unobscured Ly$\alpha$ (and UV) luminosity, which seems to correspond to a SFR of $\approx20$\,M$_{\odot}$\,yr$^{-1}$. Galaxies with higher SFRs exist in large numbers and will have higher intrinsic luminosities but dust extinction reduces the observable flux in a non-linear way, resulting in a limit to the observed luminosity. The reason why AGN become prevalent above this limit is because the physics of Ly$\alpha$ (and UV) production and escape is able to scale up much higher without dust limiting it. If this is the case, then the evolution of the AGN LAE fraction towards lower values, or towards higher Ly$\alpha$ observed luminosities, may imply that at high redshift galaxies can form stars at higher rates without dust limiting the observed Ly$\alpha$ luminosity. There is evidence for the brightest LAEs at $z\sim7$ being mergers unobscured by dust, while structures with similar Ly$\alpha$ luminosity at $z=2-3$ show heavy obscuration \citep[see e.g.][]{Matthee2019}. This might be possible under much more metal poor conditions, together with hard and intense radiation fields that limit dust production or even destroy dust very effectively, with consequences for even the escape of LyC photons at high redshift. Another competing effect may be the reduction of the number of AGN LAEs which may happen faster than the decline in the number density of SF-powered LAEs for a given observed Ly$\alpha$ luminosity. These scenarios can only be fully explored and tested with future deep rest-frame UV and optical spectroscopy of luminous LAEs at high redshift \citep[][]{Sobral2018}.

\subsection{The SFRs of LAEs}\label{SFR_LAES_results}

\subsubsection{FIR SFRs}

A total of 46 LAEs are individually detected in at least one of the FIR bands (100, 160, 250, 350 and 500\,$\mu$m) in \cite{Shuowen2018}. We use the fluxes from \cite{Shuowen2018} and estimate the associated SFR$_{\rm IR}$ following Section \ref{FIR_SFR}. The average (median) SFR$\rm _{IR}$ of FIR detected LAEs stands at $340^{+290}_{-260}$\,M$_{\odot}$\,yr$^{-1}$ ($200^{+430}_{-110}$\,M$_{\odot}$\,yr$^{-1}$). Most (31 of 46, 67\%) of the FIR-detected LAEs have SFRs of 30-300\,M$_{\odot}$\,yr$^{-1}$, with only 6 (13\%) having SFRs\,$>600$\,M$_{\odot}$\,yr$^{-1}$ and the remaining 20\% having values in between 300 and 600\,M$_{\odot}$\,yr$^{-1}$.

We also stack our LAEs in the five FIR bands and recover SFR upper-limits of 30, 45 and 300\,M$_{\odot}$\,yr$^{-1}$ for the redshift ranges of $2.2<z<2.7$, $2.7<z<3.5$ and $3.5<z<5.8$. Our direct detections stand above these limits, with average SFRs of 114, 320 and 900$\rm \,M_{\odot}\,yr^{-1}$ for the same redshift ranges. It is also worth noting that 35 out of the 46 LAEs (76\%) with at least one FIR detection are X-ray or radio AGN LAEs. In summary, most ($\sim$99\%) LAEs remain undetected in the de-blended FIR catalogue of \cite{Shuowen2018} and they also remain undetected once stacked, implying SFRs below a few tens of M$_{\odot}$\,yr$^{-1}$.

\subsubsection{Radio SFRs}

Stacking all SC4K LAEs in the 3\,GHz band (excluding radio LAEs; see Section \ref{radio_z_results}) in the radio results in an average SFR of $9.3^{+3.0}_{-2.4}$\,M$_{\odot}$\,yr$^{-1}$ and a median of $8.6^{+2.5}_{-2.0}$\,M$_{\odot}$\,yr$^{-1}$, well under the limits imposed by the FIR stacking measurements. We are able to further split the sample in redshift bins. For $2.2<z<2.7$ and $2.7<z<3.5$, our radio stacking yield average SFRs of $9.0^{+3.0}_{-2.5}$ and $10.6^{+4.1}_{-3.2}$\,M$_{\odot}$ yr$^{-1}$, respectively (see Tables \ref{table:stacksourc} and \ref{table:appendix_stacks_radio}), and corresponding median SFRs of $8.4^{+1.7}_{-1.6}$ and $9.7^{+3.4}_{-2.9}$\,M$_{\odot}$ yr$^{-1}$. For $z>3.5$ we find a mean (median) upper-limit of $<43.5$ (37.4)\,M$_{\odot}$\,yr$^{-1}$. We also obtain radio stacks in terms of Ly$\alpha$ luminosity, measuring higher radio SFRs for higher Ly$\alpha$ luminosities (see Table \ref{table:appendix_stacks_radio}).

\subsubsection{Ly$\alpha$ SFRs}

We estimate our Ly$\alpha$ SFRs (Section \ref{Line_SFRs}) by excluding AGN (radio and X-ray) from the sample, because Ly$\alpha$ emission may be coming from the accretion process of the SMBHs for these sources and result in a biased SFR measurement when not excluded (see Section \ref{Xlum_z_results}). Excluding the direct detections should reduce this problem but we caution that the absence of X-ray or radio detections does not mean there might not still be contribution from low-luminosity AGN that have remained undetected. We obtain a median SFR of $6.0^{+7.0}_{-2.7}$\,M$_{\odot}$\,yr$^{-1}$ for the entire SC4K sample. Santos et al. (in prep.) estimate the SFRs of LAEs by making use of both the recipe from \cite{sobral2019} and by using {\sc magphys} \citep[][]{daCunha2008} to obtain SED-derived SFRs, finding a median SFR$_{\rm Ly\alpha} = 5.7^{+7.0}_{-2.6}$\,M$_{\odot}$\,yr$^{-1}$ and a median SFR$\rm _{SED}=4.5^{+9.5}_{-2.6}$\,M$_{\odot}$\,yr$^{-1}$, fully consistent with our results and confirming the low SFRs of SC4K LAEs.

We also split the sample in three different redshift bins, finding similar SFRs with medians of $4.1^{+3.7}_{-1.9}$, $6.0^{+5.3}_{-2.5}$ and $9.8^{+9.7}_{-5.2}$\,M$_{\odot}$\,yr$^{-1}$ for LAEs at $2.2<z<2.7$, $2.7<z<3.5$ and $z>3.5$, respectively (see also Table \ref{table:stacksourc}).

%%%%%%%%%%%%%%
% Figure 15 - BHAR vs SFR
%%%%%%%%%%%%%%
\begin{figure}
\centering
\includegraphics[width=8.7cm]{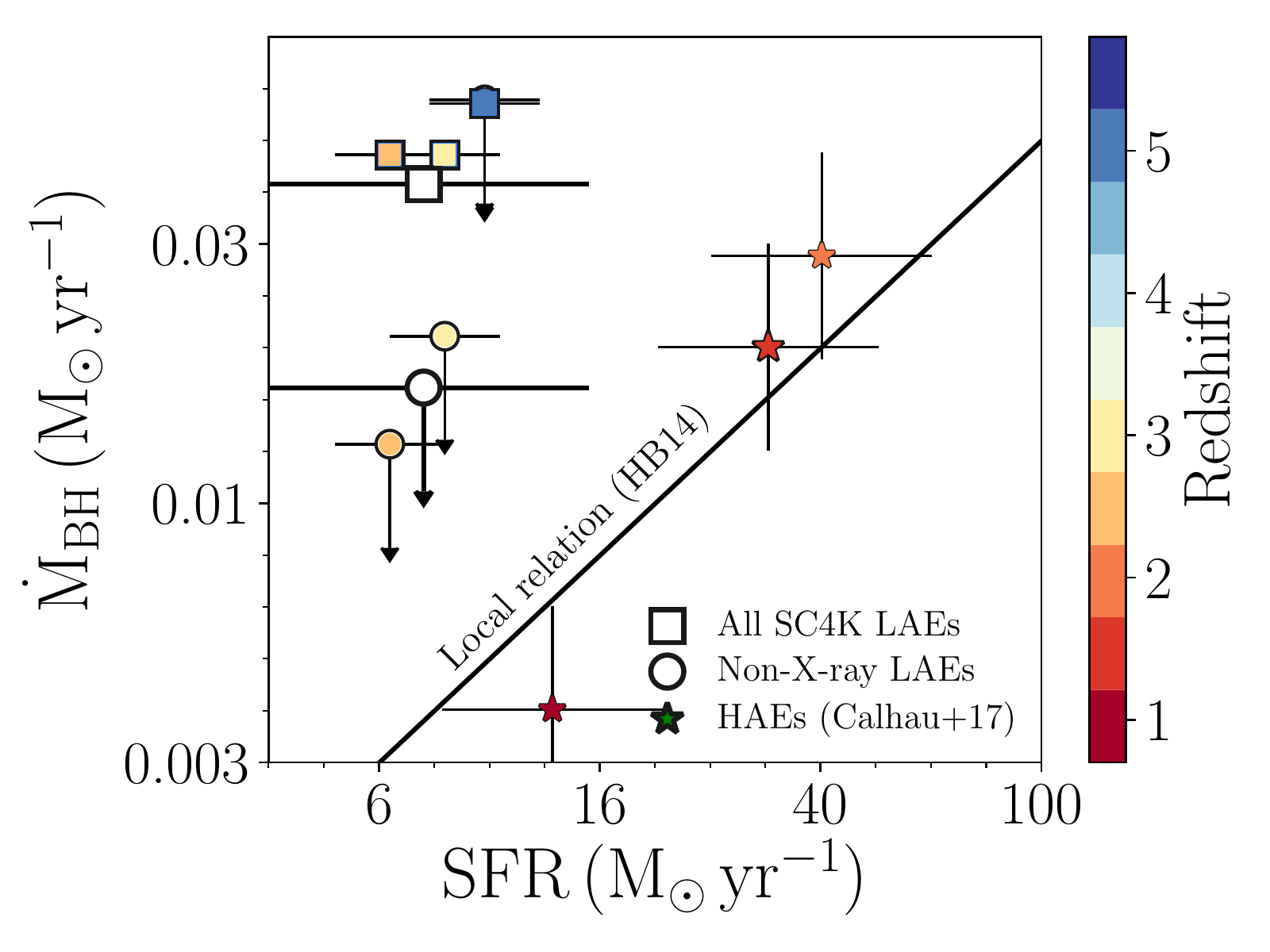}
\caption{The average star formation rate (SFR) of LAEs versus their average black hole accretion rate (BHAR or $\rm \dot{M}_{\rm BH}$). The square markers present the results including X-ray LAEs, while the circles show the results excluding those sources. The data points are coloured following the median redshift of the respective stack. The white markers are the stacks for the entire redshift range of the sample ($z=2-6$) including all LAEs (square) and excluding X-ray LAEs (circle). We compare our results with a similar analysis of H$\alpha$ selected SF galaxies from HiZELS \citep[see ][]{Calhau2017}. The black line represents the local relation between the BHAR and SFR of galaxies, taken from \citet{HeckmanBest2014}. Our results show that the non-AGN LAEs likely have BHAR/SFR ratios consistent with BH-galaxy co-evolution at much lower SFRs than the typical H$\alpha$ emitters. However, AGN LAEs are growing their super-massive black holes at significantly higher rates, many times above the local relation.}
\label{fig:SFRBHAR}
\end{figure}

\subsubsection{SFR comparison and Ly$\alpha$ escape fraction from radio SFR}\label{esc_frac_res}

The derived SFRs from different methods produce consistent results for the SC4K LAEs. In particular, the median Ly$\alpha$ SFRs and the median radio SFRs reveal excellent agreement with SFR$\rm _{Ly\alpha} = 6.0^{+7.0}_{-2.7}$\,M$_{\odot}$ yr$^{-1}$ and SFR$\rm _{1.4\,GHz}=8.6^{+2.5}_{-2.0}$\,M$_{\odot}$ yr$^{-1}$. In this work, we take the average between the Ly$\alpha$-derived values and the radio-derived values (if available) as the SFRs\footnote{The errors in the combined SFR are estimated by applying standard error propagation. For the highest redshift LAEs we use the SFR from Ly$\alpha$ only.} of LAEs. For the full sample, we find a $\rm SFR = 7.2^{+6.6}_{-2.8}\,M_{\odot}\,yr^{-1}$.

The availability of radio SFRs allows us to estimate the Ly$\alpha$ escape fraction for SC4K LAEs. We use equation \ref{eqn:SFR_fesc} using the observed median Ly$\alpha$ luminosity and obtain a Ly$\alpha$ f$\rm _{esc}=0.5 \pm 0.2$. We can also estimate f$\rm _{esc}$ with equation \ref{eqn:EW_fesc} \citep{sobral2019} and obtain f$\rm _{esc}=0.7 \pm 0.3$, showing a good agreement within the uncertainties. Our results represent the first time radio has been used to determine the escape fraction of high-$z$ LAEs and that the \cite{sobral2019} calibration can be tested with an independent method at high redshift \citep[see also][]{Santos2020}.

\subsubsection{The black hole-to-galaxy growth of LAEs}\label{BHAR/SFR_results}

Having determined the BHARs and SFRs for different sub-samples of LAEs, we can attempt to investigate the relative black hole-to-galaxy growth of LAEs and any evolution with redshift. We show the results in Figure \ref{fig:SFRBHAR}. Considering the full sample of LAEs, we find a very high BHAR for the SFR inferred. However, this is mostly because the population of LAEs is made of 1) a bulk of SF galaxies with low SFRs which dominate the numbers and SFRs and 2) a small fraction of X-ray AGN LAEs which dominate the X-ray emission and become dominant at high Ly$\alpha$ luminosities and towards $z\sim2$. Indeed, if we exclude the X-ray LAEs, our results suggest that the bulk of LAEs (without X-ray emission) are consistent with co-evolution between their super-massive black holes and their stellar populations (Figure \ref{fig:SFRBHAR}).

We can further quantify our results by computing and interpreting the BHAR/SFR ratio of each sub-population (see Table \ref{table:stacksourc}). For the entire LAE sample (including the X-ray LAEs) we find an average BHAR/SFR\,$\approx0.005$, eight times higher than what is expected for the local relation \citep[BHAR/SFR\,$\approx0.0006$, see][and Figure \ref{fig:SFRBHAR}]{HeckmanBest2014}. Excluding the X-ray LAEs from the sample results in a lower ratio of  BHAR/SFR\,$<0.0022$, only a factor 3.5 from what one would expect to establish the local relation in a co-evolution scenario between the growth of the super massive black hole and the host galaxy. 

We find that the typical BHAR/SFR for LAEs decreases with redshift by a factor of just over 3. We find BHAR/SFR\,$\approx0.007$ for $z\sim2.2-2.7$, decreasing to BHAR/SFR\,$\approx0.005$ at $z\sim2.7-3.5$ and BHAR/SFR\,$<0.006$ at $z\sim3.5-6$. When excluding X-ray LAEs we can only obtain upper limits, so we are not able to investigate any potential evolution.

In Figure \ref{fig:SFRBHAR} we also compare our results for LAEs with the results of \cite{Calhau2017} for H$\alpha$ emitters at $\rm 0.8<z<2.2$. Our LAEs have, on average, BHARs comparable to those of H$\alpha$ emitters at $\rm 0.8<z<1.5$, while the SFRs of Ly$\alpha$ emitters are around an order of magnitude smaller than that of H$\alpha$ emitters. The average relative black hole-to-galaxy growth ratio for all LAEs is much higher than for all H$\alpha$ emitters. However, when excluding X-ray LAEs the LAE population may well be fully consistent with the BHAR/SFR ratios measured for the highly star-forming H$\alpha$ selected sources (see Figure \ref{fig:SFRBHAR}), but with LAEs having significantly lower SFRs.

\section{Conclusions}\label{conclusions}

We have studied the X-ray and radio properties of 3700 LAEs (Ly$\alpha$ Emitters) at $\rm 2<z<6$ from the SC4K sample \citep[][]{SC4Kpaper}, and investigated the possible relations between those quantities and Ly$\alpha$. We made use of the publicly available data from COSMOS \citep[][]{ChandraLegacy} and use it to stack the sample and get the average X-ray luminosity and BHAR of LAEs. We also explore the radio properties of LAEs and use the available data from VLA-COSMOS \citep[][]{Smolcic2017} to also stack the sample in the radio and estimate the SFR and radio properties of typical LAEs. Our main results are:

\begin{itemize}

\item A total of 254 LAEs ($6.8\% \pm 0.4\%$) are detected by {\it Chandra} and so classified as X-ray AGN.

\item Most X-ray detections (227/254) are found at $\rm z=2.2-3.5$ with luminosities ranging from $\rm L_X = 10^{43} \, erg\, s^{-1}$ to $\rm L_X = 10^{45} \, erg \, s^{-1}$, resulting in BHARs as high as $\rm \sim4 \, M_{\odot} \, yr^{-1}$.

\item X-ray LAEs have a hardness ratio of $-0.1\pm0.2$, consistent with the global X-ray AGNs at similar redshifts. We find that about half of the X-ray LAEs have hardness ratios consistent with obscured sources, with this fraction declining with increasing X-ray luminosity, but showing no change with Ly$\alpha$ luminosity or redshift.

\item The X-ray luminosity of our LAEs correlates with the Ly$\rm \alpha$ luminosity as $\rm \log_{10}(L_X/erg\,s^{-1})=\log_{10}(L_{Ly\alpha}/erg\,s^{-1})\times(1.18 \pm 0.12)+(7.3 \pm 5.3)$, driven by the AGN present within the sample. Ly$\rm \alpha$ is likely tracing the BHAR for X-ray LAEs.

\item LAEs remain undetected in deep X-ray stacks performed by excluding X-ray LAEs (S/N\,$>3$). As a result, non X-ray LAEs present a low average BHAR of  $< 0.017$\,M$_{\odot}$\,yr$^{-1}$.

\item Overall, $3.1\% \pm 0.3\%$ (116) of our LAEs are detected in the radio, either in the 1.4\,GHz or 3\,GHz (or both) bands. 

\item We find that radio-detected LAEs have a median radio spectral index ($\alpha$) of $-1.3^{+0.4}_{-1.5}$, steeper than the global radio AGN population, which may indicate that they are good sign-posts of over-dense regions (proto-clusters) at high redshift, consistent with the clustering analysis of \cite{Khostovan2019}.

\item We find no relation between radio and Ly$\alpha$ luminosities, implying radio is tracing different processes/timescales than Ly$\alpha$ and X-rays, for the AGN LAEs.

\item The AGN fraction of LAEs increases significantly with L$_{\rm Ly\alpha}$, with the brightest LAEs being AGN dominated. The correlation is found at all redshifts, but it is found to evolve towards lower AGN fractions at higher redshift, for a fixed Ly$\alpha$ luminosity. This could be due to a shift towards higher values in the maximal observed unobscured Ly$\alpha$ luminosity, as proposed and discussed by \cite{Sobral2018}.

\item The X-ray AGN fraction drives the global AGN fraction dependence on Ly$\alpha$ luminosity. The radio AGN fraction remains relatively low with increasing Ly$\alpha$ luminosity and only grows significantly at the highest Ly$\alpha$ luminosities.

\item We are able to estimate SFRs for SC4K LAEs from radio stacking, yielding $8.6^{+2.5}_{-2.0}$\,M$_{\odot}$ yr$^{-1}$, and from Ly$\alpha$, resulting in $6.0^{+7.0}_{-2.7}$\,M$_{\odot}$\,yr$^{-1}$, fully consistent with the upper limits we obtain with FIR {\it Herschel} data.

\item We estimate the Ly$\alpha$ escape fraction of SC4K LAEs from radio SFRs (excluding AGN), obtaining $\rm f_{esc}=0.5 \pm 0.2$. We find that this is in agreement with what we obtain using \cite{sobral2019} and the Ly$\alpha$ EW$\rm _0$ ($\rm f_{esc}(EW)=0.7\pm 0.3$).

\item The full population of LAEs as a whole is growing their super-massive black holes at a relative faster rate than their host galaxies, but this is driven by a small fraction of the LAE population which is detected in the X-rays. Excluding X-ray sources, LAEs have a black hole-to-galaxy growth ratio of $\rm \log(\dot{M}_{BH}/SFR) <-2.7$, comparable to star-forming galaxies at lower redshifts and consistent with a co-evolution between their super-massive black holes and their host galaxies.
\end{itemize}

Our results reveal that LAEs at high redshift are mostly star-forming galaxies with relatively low median SFRs (7.2$\rm ^{+6.6}_{-2.8}$\,M$_{\odot}$ yr$^{-1}$) and low AGN activity (BHAR$\rm <$0.017\,M$_{\odot}$ yr$^{-1}$), but with a few ($\rm 6.8\% \pm 0.4\%$) X-ray-bright AGN where the Ly$\alpha$ emission likely comes from the accretion of matter into the central super-massive black hole. Our results therefore suggest LAEs make up a heterogeneous population of largely star-forming galaxies and a smaller number of AGN, where Ly$\alpha$ becomes an accretion indicator. The X-ray LAEs become the dominant population among LAEs at the highest Ly$\alpha$ luminosities, but there seems to be an important negative evolution of such population towards high redshift. Radio LAEs are not fully representative of the radio-selected AGN at similar redshifts. LAEs detected in the radio show properties consistent with residing in over-dense regions, but there is a general lack of correlation between the radio and Ly$\alpha$, unlike the strong Ly$\alpha$-X-ray correlations, likely due to those luminosities tracing very different timescales and consistent with significant AGN variability for AGN LAEs. Future studies are required to conduct deep spectroscopic observations of LAEs to unveil even lower BHARs, to establish the redshift evolution even more conclusively, and to identify the physical origins and their consequences for how early galaxies form and evolve.

\section*{Acknowledgments}

We thank the anonymous referee for the valuable feedback which greatly improved the paper. JC and SS acknowledge Lancaster University PhD studentships. APA acknowledges support from FCT through the project PTDC/FIS-AST/31546/2017. AS acknowledges support through a Clay Fellowship administered by the Smithsonian Astrophysical Observatory. The authors thank Richard Bower and Nicholas Ross for interesting discussions during the DEX XIV conference. The authors would also like to thank Francesca Civano, Francesca Fornasini and the \textit{Chandra} team for several comments, suggestions and interesting discussions and for making the X-ray data and exposure maps fully available for this work. This research has made use of NASA's Astrophysics Data System. This work has made use of the analysis software {\sc topcat} \citep[][]{topcat} and the public programming language \textsc{python} with the following packages: {\sc numpy} \citep[][]{numpy}, {\sc matplotlib} \citep[][]{matplotlib}, {\sc pyfits} and {\sc astropy} \citep[][]{Astropy}. This paper is based on the \href{https://youtu.be/tKb2osj3nkU}{public SC4K} sample \citep[][]{SC4Kpaper}.

\bibliographystyle{mnras}
\bibliography{BHGalaxyGrowth}

\appendix

\section{Comparison between this work and Civano et al (2016)'s results}\label{Comparing_with_Civano_and_5sigma_cuts}

\begin{figure}
	\centering
	\includegraphics[width=8.1cm]{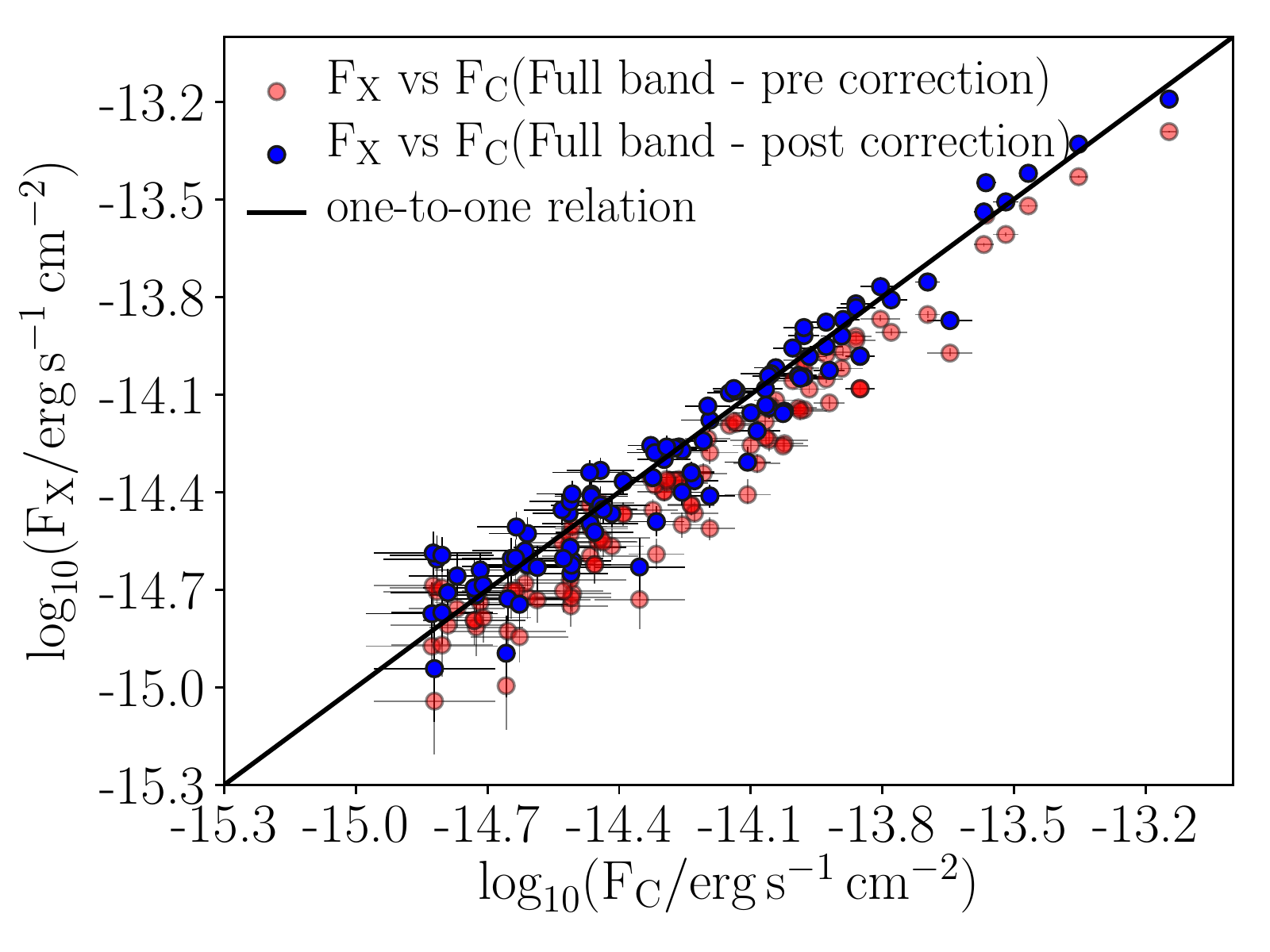}
	\caption{Comparison between the X-ray fluxes obtained in this work (F$\rm _X$) and the ones reported by \citet{ChandraLegacy} (F$\rm _C$), for the full band (0.5-7\,keV) of \textit{Chandra}. As expected, our initial aperture fluxes consistently underestimate the X-ray fluxes. After applying a median correction, we successfully recover the fluxes of \citet{ChandraLegacy} on average.}
	\label{fig:FluxesComp}
\end{figure}

In order to assure we obtain full X-ray fluxes in individual detections and for stacks we derive and apply an aperture correction, as explained in Section \ref{X-rays}.
Figure \ref{fig:FluxesComp} shows the comparison between our initial aperture fluxes and the aperture corrected ones when compared to \cite{ChandraLegacy}, showing an excellent agreement.

%%%%%%%%%%%%%%
% Figure B1- Flux comparisons
%%%%%%%%%%%%%%

Using just the \cite{ChandraLegacy}'s catalogue would be an alternative of looking at the questions explored by this work (which we also take), and we show that our results are unchanged in a qualitative way. However, since in this study we have a pre-selected sample of sources (LAEs), and because stacking is crucial to try to unveil any fainter X-ray emission statistically, it is crucial that the individual detections and the stacking methodology are self-consistent. In our analysis we make sure that i) we can reproduce the robust fluxes of \cite{ChandraLegacy} and ii) we apply a methodology that is easily transferrable to our stacking analysis in a self-consistent way and that allows us to go to a lower S/N.

\section{Comparison between this work and VLA's Bondi et al (2008) and Smol\u{c}i\'{c} et al.(2017)'s results}\label{Comparing_with_VLA}

In order to assure we obtain full radio fluxes in individual detections and for stacks we derive and apply a flux correction based on fluxes published by \cite{Schinnerer2007,Bondi2008,Smolcic2017}; see Section \ref{radio}. Figure \ref{fig:FluxesComp} shows the comparison between our initial radio fluxes and the corrected ones when compared to \cite{Smolcic2017}.

%%%%%%%%%%%%%%
% Figure C1 - radio Flux corrections
%%%%%%%%%%%%%%
\begin{figure}
	\centering
	\includegraphics[width=8.1cm]{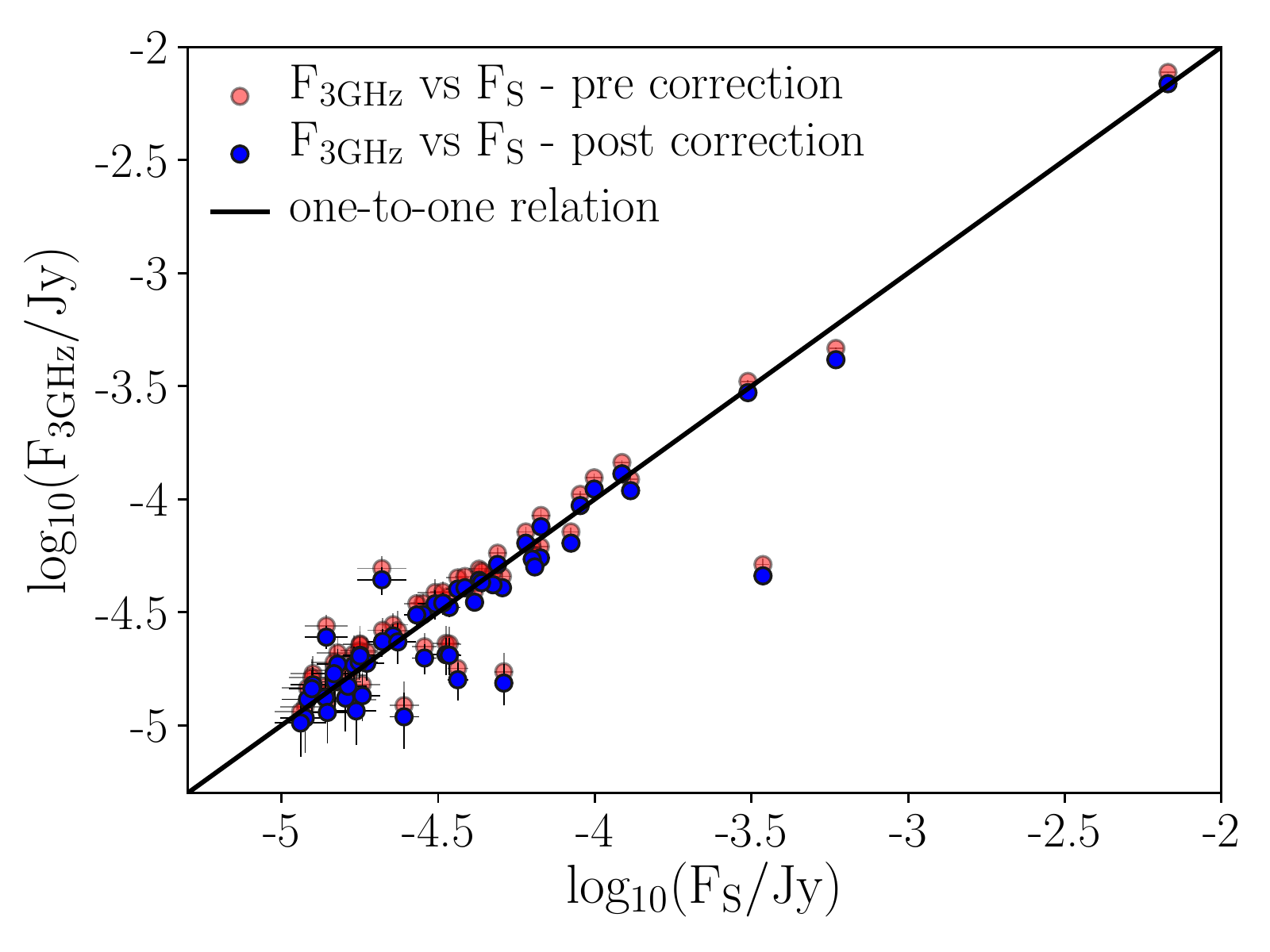}
	\caption{Comparison between our initial radio fluxes (F$\rm _{3\,GHz}$) and those of \citet{Smolcic2017} (F$\rm _S$). The black line represents a perfect agreement between measurements. The blue points show the flux difference between the two data sets after applying the correction.}
	\label{fig:radioFluxesdelta}
\end{figure}

\section{Public Catalogue of SC4K LAEs and Stacking tables}\label{AGN_catalogue}

Table \ref{table:detsourc} shows the first 3 entries in the catalogue of all SC4K LAEs which we make available with this paper. The complete catalogue includes all LAEs in the study, with X-ray and radio measurements for the LAEs detected in these bands, as well as upper limits for the ones that remain undetected, following the methods described in this study. We also include the measurements from the catalogues from \cite{ChandraLegacy}, \cite{Smolcic2017} and FIR fluxes and FIR-derived SFRs from \cite{Shuowen2018}.

We provide, in tables \ref{table:appendix_stacks_radio} and \ref{table:appendix_stacks}, extended stacking results in the X-ray and radio done in this work. All quantities are estimated following the procedures detailed in Sections \ref{X-rays} (for the X-ray analysis) and \ref{radio} (for the radio analysis). The tables are available online.

%
% Table A1 - Properties of the direct detections
%
\begin{table*}
	\centering
	\caption[]{The first 3 entries of LAEs in our SC4K \citep[adding to][]{SC4Kpaper}, fully available online. The uncertainties in the redshifts are taken from the NB or MB filter widths \citep[see][]{SC4Kpaper}. The X-ray luminosities were estimated from the full band fluxes extracted from the {\it Chandra} Legacy survey images ($\rm 0.5 - 10 \, keV$). We also show the BHARs as determined from the X-ray luminosities of the sources. The full public catalogue includes all LAEs in the study, detected and undetected in the X-rays, as well as radio, FIR and Ly$\alpha$ quantities. It also includes the measurements from the catalogues from \citet{ChandraLegacy}, \citet{Smolcic2017} and \citet{Shuowen2018}.}
	\begin{tabular}{@{}lcccccccc@{}}
	\hline
	Source ID &R.A.&Dec.& z & $\log_{10}$ L$\rm _{\rm Ly\alpha}$ & $\rm \log_{10} flux_X$ &$\log_{10}$ L$_X$& $\rm \dot{M}_{\rm BH}$ & (And 72 other\\
	&(J2000) &(J2000)& (MB/NB) & [erg\,s$^{-1}$] & [erg\,s$^{-1} \, cm^{-2}$] &[erg\,s$^{-1}$] &[M$_{\odot}$\,yr$^{-1}$] & columns)\\
	\hline
	SC4K-NB392-2 & 10:02:42.319 & +02:35:29.47 & 2.22 &42.70$^{+0.04}_{-0.06}$& $<$-14.99& $<$43.39 & $<$0.088& -\\
	SC4K-NB392-6 & 10:02:36.830 & +02:39:47.33 & 2.22 &42.60$^{+0.05}_{-0.06}$& $<$-15.00& $<$43.38 & $<$0.086&-\\
	SC4K-NB392-7 & 10:02:36.771 & +02:32:25.79 & 2.22 &42.77$^{+0.03}_{-0.05}$& -14.89$^{+0.11}_{-0.13}$ & 43.49$^{+0.11}_{-0.13}$ & 0.109$^{+0.030}_{-0.029}$ &- \\
	\hline
		\label{table:detsourc}
	\end{tabular}
\end{table*}

%%%%%%%%%%%%%%
% Table D2 - PLACEHOLDER FOR STACKS TABLE
%%%%%%%%%%%%%%
\begin{table*}
	\centering
	\caption[]{The results for the stacking analysis of the radio emission of LAEs. We determine the fluxes following the process detailed in Section \ref{radio}, using a circular aperture of 5.3$''$ and  1.2$''$ for the 1.4\,GHz and 3\,GHz, respectively. In order to make them comparable to \protect{\cite{Smolcic2017}}, we follow Section \ref{radio_alpha}. A stack is considered as having a detection when the S/N$\geq$3. For stacks with lower S/N, we provide the 3$\sigma$ upper limit.}
\begin{tabular}{@{}cccccccccc@{}}
	 \hline
	 Subsample & \# Sources & Redshift & $\rm \log_{10}$ L & $\rm \log_{10}$ L & SFR & S/N$_{\rm 14}$ & $\rm \log_{10}$ L & SFR & S/N$_{\rm 3}$\\
	 Stacked  & & (median) & Ly$\alpha$& (1.4\,GHz) & (1.4\,GHz) & & (3.0\,GHz)& (3.0\,GHz)\\
	  & & &[erg\,s$^{-1}$] &[W Hz $^{-1}$]& [M$_{\odot}$\,yr$^{-1}$]& &[W Hz $^{-1}$]& [M$_{\odot}$\,yr$^{-1}$]& \\
	 \hline
	 \hline
	 Including radio LAEs &&&&&&&\\
	 \hline
	 \hline
	Full Sample &3696&3.1$^{+1.0}_{-0.6}$& 42.9$^{+0.3}_{-0.2}$ & 23.59$^{+0.08}_{-0.07}$ & - & 6.7&23.53$^{+0.01}_{-0.01}$ & - & 50.9\\
	No X-ray LAEs &3442&3.2$^{+1.0}_{-0.6}$& 42.8$^{+0.3}_{-0.2}$ &$<$23.5 & - & $<$3 &22.70$^{+0.08}_{-0.07}$ & - & 6.5 \\
	2.2$<$z$<$2.7 &849&2.5$^{+0.0}_{-0.3}$& 42.6$^{+0.2}_{-0.1}$ & 24.00$^{+0.02}_{-0.02}$ & - & 19.6&23.81$^{+0.01}_{-0.01}$ & - & 75.6\\
	2.7$<$z$<$3.5 &2085&3.2$^{+0.2}_{-0.2}$& 42.9$^{+0.2}_{-0.1}$ &$<$23.6 & - & $<$3 &23.22$^{+0.04}_{-0.03}$ & - & 16.5 \\
	3.5$<$z$<$6 &762&4.8$^{+0.9}_{-0.7}$& 43.1$^{+0.2}_{-0.3}$ & $<$24.0 & - & $<$3 & $<$23.2 & - & $<$3\\
	42.2$<$ $\log_{10}$(L$_{\rm Ly \alpha}$)$<$43.0 &2654&3.0$^{+0.4}_{-0.5}$& 42.8$^{+0.1}_{-0.2}$ & 23.51$^{+0.08}_{-0.07}$ & - & 6.8&23.49$^{+0.01}_{-0.01}$ & - & 43.3\\
	43.0$<\rm \log_{10}(L_{ \rm Ly\alpha})<$43.3 &770&3.3$^{+1.5}_{-0.4}$& 43.1$^{+0.1}_{-0.1}$ & 23.53$^{+0.13}_{-0.16}$ & - & 3.2&23.34$^{+0.04}_{-0.03}$ & - & 13.4\\
	 43.3$<\rm \log_{10}(L_{\rm Ly\alpha})<$43.5 &157&4.1$^{+1.2}_{-1.2}$& 43.4$^{+0.1}_{-0.0}$ & 24.27$^{+0.10}_{-0.07}$ & - & 6.5&24.14$^{+0.02}_{-0.02}$ & - & 23.5\\
	 43.5$<\rm \log_{10}(L_{\rm Ly\alpha})<$43.8 &82&3.3$^{+2.0}_{-0.4}$& 43.6$^{+0.1}_{-0.1}$ & 24.16$^{+0.32}_{-0.13}$ & - & 3.8&23.96$^{+0.04}_{-0.03}$ & - & 16.4\\
	43.8$<\rm \log_{10}(L_{\rm Ly\alpha})<$44.8 &32&3.1$^{+0.2}_{-0.6}$& 43.9$^{+0.2}_{-0.1}$ & 24.28$^{+0.11}_{-0.12}$ & - & 4.2&23.95$^{+0.04}_{-0.04}$ & - & 11.4\\
	 \hline
	 \hline
	 Excluding radio LAEs &&&&&&&\\
	 \hline
	 \hline
	Full Sample &3576&3.2$^{+1.0}_{-0.6}$& 42.8$^{+0.3}_{-0.2}$ &$<$23.4 &$<$84.0 & $<$3 &22.47$^{+0.12}_{-0.13}$ & 9.3$^{+3.0}_{-2.4}$& 3.9 \\
	2.2$<$z$<$2.7 &817&2.5$^{+0.0}_{-0.3}$& 42.6$^{+0.2}_{-0.1}$ &$<$23.2 &$<$51.6 & $<$3 &22.45$^{+0.12}_{-0.14}$ & 9.0$^{+3.0}_{-2.5}$& 3.6 \\
	2.2$<$z$<$3.5 &2013&3.2$^{+0.2}_{-0.2}$& 42.9$^{+0.2}_{-0.1}$ &$<$23.5 &$<$110.6 & $<$3 &22.52$^{+0.14}_{-0.16}$ & 10.6$^{+4.1}_{-3.2}$& 3.3 \\
	3.5$<$z$<$6 &746&4.8$^{+0.9}_{-0.7}$& 43.1$^{+0.2}_{-0.3}$ & $<$24.0 &$<$288.9& $<$3 & $<$23.1 & $<$43.5& $<$3\\
	42.2$<$ $\log_{10}$(L$_{\rm Ly \alpha}$)$<$43.0 &2601&3.0$^{+0.4}_{-0.5}$& 42.8$^{+0.1}_{-0.2}$ & $<$23.3 &$<$66.3& $<$3 & $<$22.4 & $<$8.2& $<$3\\
	43.0$<\rm \log_{10}(L_{ \rm Ly\alpha})<$43.3 &738&3.3$^{+1.5}_{-0.4}$& 43.1$^{+0.1}_{-0.1}$ &$<$23.6 &$<$126.2 & $<$3 &22.73$^{+0.14}_{-0.17}$ & 16.9$^{+6.3}_{-5.5}$& 3.0 \\
	 43.3$<\rm \log_{10}(L_{\rm Ly\alpha})<$43.5 &143&4.6$^{+0.7}_{-1.6}$& 43.4$^{+0.1}_{-0.0}$ & $<$24.3 &$<$673.6& $<$3 & $<$23.4 & $<$77.2& $<$3\\
	 43.5$<\rm \log_{10}(L_{\rm Ly\alpha})<$43.8 &68&4.6$^{+1.1}_{-1.6}$& 43.6$^{+0.1}_{-0.1}$ &$<$25.0 &$<$3540.2 & $<$3 &23.75$^{+0.15}_{-0.12}$ & 177.7$^{+72.2}_{-41.4}$& 4.3 \\
	43.8$<\rm \log_{10}(L_{\rm Ly\alpha})<$44.8 &25&3.2$^{+0.2}_{-0.6}$& 43.9$^{+0.2}_{-0.1}$ &$<$24.3 &$<$640.5 & $<$3 &23.58$^{+0.11}_{-0.14}$ & 120.5$^{+33.3}_{-33.7}$& 3.6 \\
	 \hline
	 \label{table:appendix_stacks_radio}
	\end{tabular}
\end{table*}

%%%%%%%%%%%%%%
% Table D1 - PLACEHOLDER FOR STACKS TABLE
%%%%%%%%%%%%%%
\begin{table*}
	\centering
	\caption[]{The results of our stacking analysis of the X-ray emission of LAEs. The redshift was estimated by taking the median and using the 16th and 84th percentiles as the errors. We use this median redshift to determine the luminosity distances used in calculating the x-ray luminosities. The fluxes were estimated using a circular aperture of 7.9$''$. We apply further correction factors, as detailed in Section \ref{X-rays}, including an aperture correction. A stack is considered to have a detection if the S/N$\geq$3 and we provide the 3$\sigma$ upper-limit in the case of a non-detection.}
 \begin{tabular}{@{}cccccccc@{}}
	 \hline
	 Subsample & \# Sources & Redshift & $\rm \log_{10}$ L &$\rm \log_{10}$ F & $\rm \log_{10}$ L & BHAR & S/N\\
	 Stacked  & & (median) & Ly$\alpha$& 0.5 - 7 keV & X-rays &0.5 - 7 keV &0.5 - 7 keV \\
	  & & &[erg\,s$^{-1}$]&[erg\,s$^{-1}$\,cm$^{-2}$] &[erg\,s$^{-1}$] &[M$_{\odot}$\,yr$^{-1}$]& \\
	 \hline
	 \hline
	 Including X-ray LAEs &&&&&&&\\
	 \hline
	 \hline
	Full Sample &3700&3.1$^{+1.0}_{-0.6}$& 42.85$^{+0.29}_{-0.17}$ & -15.61$^{+0.06}_{-0.07}$ &43.06$^{+0.06}_{-0.07}$ & 0.041$^{+0.006}_{-0.006}$& 6.7\\
	X-ray LAEs only &254&3.0$^{+0.4}_{-0.5}$& 43.08$^{+0.61}_{-0.31}$ & -14.33$^{+0.01}_{-0.01}$ &44.31$^{+0.01}_{-0.01}$ & 0.720$^{+0.015}_{-0.011}$& 65.8\\
	2.2$<$z$<$2.7 &849&2.5$^{+0.0}_{-0.3}$& 42.64$^{+0.24}_{-0.15}$ & -15.37$^{+0.05}_{-0.05}$ &43.12$^{+0.05}_{-0.05}$ & 0.047$^{+0.005}_{-0.005}$& 9.1\\
	2.7$<$z$<$3.5 &2085&3.2$^{+0.2}_{-0.2}$& 42.86$^{+0.22}_{-0.12}$ & -15.56$^{+0.07}_{-0.07}$ &43.12$^{+0.07}_{-0.07}$ & 0.047$^{+0.008}_{-0.007}$& 6.4\\
	3.5$<$z$<$6 &766&4.8$^{+0.9}_{-0.7}$& 43.07$^{+0.24}_{-0.28}$ &$<$-15.8 &$<$43.2 & $<$0.059& $<$3\\
	42.2$<$ $\log_{10}$(L$_{\rm Ly \alpha}$)$<$42.6 &384&2.5$^{+0.0}_{-0.3}$& 42.52$^{+0.05}_{-0.07}$ &$<$-15.7 &$<$42.7 & $<$0.020& $<$3\\
	42.6$<\rm \log_{10}(L_{ \rm Ly\alpha})<$42.7 &323&2.5$^{+0.8}_{-0.0}$& 42.67$^{+0.02}_{-0.04}$ &$<$-15.8 &$<$42.7 & $<$0.017& $<$3\\
	 42.7$<\rm \log_{10}(L_{\rm Ly\alpha})<$42.8 &762&3.0$^{+0.4}_{-0.5}$& 42.75$^{+0.03}_{-0.03}$ &$<$-15.9 &$<$42.7 & $<$0.019& $<$3\\
	 42.8$<\rm \log_{10}(L_{\rm Ly\alpha})<$42.9 &686&3.2$^{+0.2}_{-0.3}$& 42.85$^{+0.03}_{-0.03}$ &$<$-15.7 &$<$43.0 & $<$0.032& $<$3\\
	42.9$<\rm \log_{10}(L_{\rm Ly\alpha})<$43.0 &500&3.2$^{+1.0}_{-0.3}$& 42.95$^{+0.03}_{-0.03}$ & -15.53$^{+0.10}_{-0.08}$ &43.15$^{+0.10}_{-0.08}$ & 0.050$^{+0.013}_{-0.009}$& 5.7\\
	43.0$<\rm \log_{10}(L_{\rm Ly\alpha})<$43.1 &334&3.2$^{+1.0}_{-0.3}$& 43.05$^{+0.03}_{-0.03}$ & -15.74$^{+0.15}_{-0.15}$ &42.94$^{+0.15}_{-0.15}$ & 0.031$^{+0.013}_{-0.009}$& 3.5\\
	43.1$<\rm \log_{10}(L_{\rm Ly\alpha})<$43.2 &242&3.3$^{+1.5}_{-0.5}$& 43.15$^{+0.04}_{-0.04}$ & -15.38$^{+0.06}_{-0.06}$ &43.35$^{+0.06}_{-0.06}$ & 0.080$^{+0.013}_{-0.011}$& 7.6\\
	43.2$<\rm \log_{10}(L_{\rm Ly\alpha})<$43.4 &302&4.1$^{+0.9}_{-1.2}$& 43.27$^{+0.07}_{-0.05}$ & -15.32$^{+0.07}_{-0.05}$ &43.59$^{+0.07}_{-0.05}$ & 0.138$^{+0.024}_{-0.015}$& 9.2\\
	43.4$<\rm \log_{10}(L_{\rm Ly\alpha})<$43.8 &134&3.5$^{+2.1}_{-0.5}$& 43.53$^{+0.16}_{-0.09}$ & -14.71$^{+0.02}_{-0.02}$ &44.07$^{+0.02}_{-0.02}$ & 0.416$^{+0.022}_{-0.017}$& 24.3\\
	43.8$<\rm \log_{10}(L_{\rm Ly\alpha})<$44.8 &32&3.1$^{+0.2}_{-0.6}$& 43.94$^{+0.20}_{-0.09}$ & -14.22$^{+0.01}_{-0.01}$ &44.46$^{+0.01}_{-0.01}$ & 1.018$^{+0.019}_{-0.022}$& 47.3\\
	 \hline
	 \hline
	 Excluding Civano$+$16 LAEs &&&&&&&\\
	 \hline
	 \hline
	Full Sample &3600&3.2$^{+1.0}_{-0.6}$& 42.85$^{+0.27}_{-0.17}$ &$<$-16.0 &$<$42.7 & $<$0.016& $<$3\\
	2.2$<$z$<$2.7 &816&2.5$^{+0.0}_{-0.0}$& 42.63$^{+0.22}_{-0.14}$ &$<$-15.9 &$<$42.6 & $<$0.014& $<$3\\
	2.7$<$z$<$3.5 &2021&3.2$^{+0.2}_{-0.2}$& 42.86$^{+0.21}_{-0.12}$ &$<$-15.9 &$<$42.8 & $<$0.020& $<$3\\
	3.5$<$z$<$6 &763&4.8$^{+0.9}_{-0.7}$& 43.07$^{+0.24}_{-0.28}$ &$<$-15.8 &$<$43.3 & $<$0.064& $<$3\\
	42.2$<$ $\log_{10}$(L$_{\rm Ly \alpha}$)$<$42.6 &382&2.5$^{+0.0}_{-0.3}$& 42.51$^{+0.05}_{-0.07}$ &$<$-15.8 &$<$42.7 & $<$0.019& $<$3\\
	42.6$<\rm \log_{10}(L_{ \rm Ly\alpha})<$42.7 &321&2.5$^{+0.8}_{-0.0}$& 42.67$^{+0.02}_{-0.04}$ &$<$-15.8 &$<$42.7 & $<$0.017& $<$3\\
	 42.7$<\rm \log_{10}(L_{\rm Ly\alpha})<$42.8 &759&3.0$^{+0.4}_{-0.4}$& 42.75$^{+0.03}_{-0.03}$ &$<$-15.9 &$<$42.7 & $<$0.019& $<$3\\
	 42.8$<\rm \log_{10}(L_{\rm Ly\alpha})<$42.9 &683&3.2$^{+0.2}_{-0.3}$& 42.85$^{+0.03}_{-0.03}$ &$<$-15.7 &$<$43.0 & $<$0.033& $<$3\\
	42.9$<\rm \log_{10}(L_{\rm Ly\alpha})<$43.0 &489&3.2$^{+1.0}_{-0.3}$& 42.95$^{+0.03}_{-0.04}$ &$<$-15.7 &$<$43.0 & $<$0.033& $<$3\\
	43.0$<\rm \log_{10}(L_{\rm Ly\alpha})<$43.1 &327&3.2$^{+1.0}_{-0.3}$& 43.05$^{+0.03}_{-0.03}$ &$<$-15.7 &$<$43.0 & $<$0.034& $<$3\\
	43.1$<\rm \log_{10}(L_{\rm Ly\alpha})<$43.2 &238&3.3$^{+1.5}_{-0.4}$& 43.15$^{+0.04}_{-0.04}$ &$<$-15.8 &$<$43.0 & $<$0.033& $<$3\\
	43.2$<\rm \log_{10}(L_{\rm Ly\alpha})<$43.4 &280&4.6$^{+0.5}_{-1.6}$& 43.27$^{+0.07}_{-0.05}$ &$<$-15.7 &$<$43.3 & $<$0.075& $<$3\\
	43.4$<\rm \log_{10}(L_{\rm Ly\alpha})<$43.8 &107&4.6$^{+1.2}_{-1.6}$& 43.51$^{+0.12}_{-0.08}$ & -15.43$^{+0.11}_{-0.11}$ &43.57$^{+0.11}_{-0.11}$ & 0.131$^{+0.039}_{-0.029}$& 4.6\\
	43.8$<\rm \log_{10}(L_{\rm Ly\alpha})<$44.8 &13&3.3$^{+1.8}_{-0.2}$& 43.96$^{+0.31}_{-0.10}$ & -14.64$^{+0.03}_{-0.03}$ &44.09$^{+0.03}_{-0.03}$ & 0.434$^{+0.028}_{-0.032}$& 13.5\\
	 \hline
	 \hline
	 \hline
	 Excluding X-ray LAEs &&&&&&&\\
	 \hline
	 \hline
	Full Sample &3446&3.2$^{+1.0}_{-0.6}$& 42.85$^{+0.27}_{-0.17}$ &$<$-16.0 &$<$42.7 & $<$0.017& $<$3\\
	2.2$<$z$<$2.7 &787&2.5$^{+0.0}_{-0.0}$& 42.63$^{+0.22}_{-0.14}$ &$<$-15.9 &$<$42.6 & $<$0.013& $<$3\\
	2.2$<$z$<$3.5 &1920&3.2$^{+0.2}_{-0.2}$& 42.86$^{+0.20}_{-0.12}$ &$<$-15.9 &$<$42.8 & $<$0.021& $<$3\\
	3.5$<$z$<$6 &739&4.8$^{+0.9}_{-0.7}$& 43.07$^{+0.23}_{-0.28}$ &$<$-15.8 &$<$43.2 & $<$0.060& $<$3\\
	42.2$<$ $\log_{10}$(L$_{\rm Ly \alpha}$)$<$42.6 &370&2.5$^{+0.0}_{-0.3}$& 42.52$^{+0.05}_{-0.07}$ &$<$-15.8 &$<$42.7 & $<$0.018& $<$3\\
	42.6$<\rm \log_{10}(L_{ \rm Ly\alpha})<$42.7 &315&2.5$^{+0.8}_{-0.0}$& 42.67$^{+0.02}_{-0.04}$ &$<$-15.9 &$<$42.6 & $<$0.015& $<$3\\
	 42.7$<\rm \log_{10}(L_{\rm Ly\alpha})<$42.8 &734&3.0$^{+0.4}_{-0.4}$& 42.75$^{+0.03}_{-0.03}$ &$<$-15.9 &$<$42.7 & $<$0.020& $<$3\\
	 42.8$<\rm \log_{10}(L_{\rm Ly\alpha})<$42.9 &656&3.2$^{+0.2}_{-0.3}$& 42.85$^{+0.03}_{-0.03}$ &$<$-15.7 &$<$43.0 & $<$0.032& $<$3\\
	42.9$<\rm \log_{10}(L_{\rm Ly\alpha})<$43.0 &470&3.2$^{+1.0}_{-0.3}$& 42.95$^{+0.03}_{-0.04}$ &$<$-15.7 &$<$43.0 & $<$0.033& $<$3\\
	43.0$<\rm \log_{10}(L_{\rm Ly\alpha})<$43.1 &312&3.2$^{+1.0}_{-0.3}$& 43.05$^{+0.03}_{-0.03}$ &$<$-15.8 &$<$42.9 & $<$0.029& $<$3\\
	43.1$<\rm \log_{10}(L_{\rm Ly\alpha})<$43.2 &223&3.3$^{+1.6}_{-0.4}$& 43.15$^{+0.04}_{-0.04}$ &$<$-15.8 &$<$43.0 & $<$0.032& $<$3\\
	43.2$<\rm \log_{10}(L_{\rm Ly\alpha})<$43.4 &268&4.6$^{+0.5}_{-1.6}$& 43.27$^{+0.07}_{-0.05}$ &$<$-15.7 &$<$43.3 & $<$0.076& $<$3\\
	43.4$<\rm \log_{10}(L_{\rm Ly\alpha})<$43.8 &91&4.8$^{+1.0}_{-1.9}$& 43.51$^{+0.10}_{-0.08}$ &$<$-15.5 &$<$43.5 & $<$0.123& $<$3\\
	43.8$<\rm \log_{10}(L_{\rm Ly\alpha})<$44.8 &6&3.2$^{+2.0}_{-0.4}$& 43.88$^{+0.14}_{-0.03}$ & -15.22$^{+0.15}_{-0.22}$ &43.48$^{+0.15}_{-0.22}$ & $<$0.109& 2.1 \\
	 \hline
	 \hline
	 \label{table:appendix_stacks}
	\end{tabular}
\end{table*}

\bsp
\label{lastpage}
\end{document}